\documentclass[9pt,shortpaper,twoside,web]{ieeecolor}
\usepackage{generic}
\usepackage{cite}
\usepackage{amsmath,amssymb,amsfonts}
\usepackage{algorithmic}
\usepackage{graphicx}
\usepackage{textcomp}
\usepackage[table,xcdraw]{xcolor}
\usepackage[acronym]{glossaries}
\usepackage{adjustbox}
\usepackage{svg}
\usepackage{makecell} 

\usepackage[labelformat=empty]{subfig}
\usepackage[margin=1in]{geometry}
\usepackage{pdfpages}

\usepackage{acronym}
\usepackage{siunitx}
\usepackage{bm}

\usepackage[T1]{fontenc}
\usepackage{times}

\let\labelindent\relax
\usepackage{enumitem}
\usepackage{pifont}

\usepackage{siunitx}
\DeclareSIUnit\mmHg{mmHg}

\usepackage{pifont}
\newcommand{\cmark}{\ding{51}} 
\newcommand{\xmark}{\ding{55}} 

\usepackage{makecell}

\usepackage{array}

\makeglossaries

\usepackage{comment}

\usepackage{nohyperref}

\definecolor{matplotlib0}{HTML}{1f77b4}
\definecolor{matplotlib1}{HTML}{d62728}
\definecolor{matplotlib2}{HTML}{2ca02c}
\definecolor{matplotlib3}{HTML}{ff7f0e}
\definecolor{matplotlib4}{HTML}{9467bd}
\definecolor{matplotlib5}{HTML}{8c564b}
\definecolor{matplotlib6}{HTML}{e377c2}
\definecolor{matplotlib7}{HTML}{7f7f7f}
\definecolor{matplotlib8}{HTML}{bcbd22}
\definecolor{matplotlib9}{HTML}{17becf}

\usepackage{mathtools} 

\usepackage{booktabs}
\usepackage{multirow}
\usepackage{colortbl}
\usepackage{tablefootnote}
\usepackage{threeparttable}
\usepackage{array}
\usepackage{xcolor}

\usepackage{tikz}
\usetikzlibrary{shapes,arrows,positioning}

\usepackage{pgfplots}
\definecolor{color0}{rgb}{0.12156862745098,0.466666666666667,0.705882352941177} 
\definecolor{color1}{rgb}{1,0.498039215686275,0.0549019607843137}
\definecolor{color2}{rgb}{0.172549019607843,0.627450980392157,0.172549019607843} 
\definecolor{color3}{rgb}{0.83921568627451,0.152941176470588,0.156862745098039} 
\definecolor{color4}{rgb}{0.580392156862745,0.403921568627451,0.741176470588235}
\definecolor{colorblue}{rgb}{0.12156862745098,0.466666666666667,0.705882352941177} 
\definecolor{colorgreen}{rgb}{0.172549019607843,0.627450980392157,0.172549019607843} 
\definecolor{colorred}{rgb}{0.83921568627451,0.152941176470588,0.156862745098039} 
\definecolor{colorblack}{rgb}{0,0,0} 
\definecolor{amber}{rgb}{1.0, 0.75, 0.0}
\definecolor{cyan(process)}{rgb}{0.0, 0.72, 0.92}
\definecolor{colororange}{rgb}{1,0.56,0} 
\definecolor{colorbrown}{rgb}{0.62, 0.42, 0.21} 
\definecolor{colorpurple}{rgb}{0.33, 0.033, 0.51}
\usepgfplotslibrary{fillbetween}
\usepgfplotslibrary{colormaps}
\pgfplotsset{compat=1.16}

\pgfplotscreateplotcyclelist{matplotlib}{
  {matplotlib0},
  {matplotlib1},
  {matplotlib2},
  {matplotlib3},
  {matplotlib4},
  {matplotlib5},
  {matplotlib6},
  {matplotlib7},
  {matplotlib8},
  {matplotlib9}
}
\pgfplotsset{every axis/.append style={
    cycle list name=matplotlib,
}}
\pgfdeclareplotmark{mystar}{
    \node[star,star point ratio=2.25,minimum size=6pt,
          inner sep=0pt,draw=black,solid,fill=red] {};
}

\usepackage{listings}
\usepgfplotslibrary{groupplots}

\definecolor{code_default}{HTML}{000000}
\definecolor{code_keyword}{HTML}{AC4142}
\definecolor{code_identifier}{HTML}{D28445}

\lstdefinelanguage{RISCV}{
  sensitive=false,
  morecomment=[l]{//},
  alsoletter={.},
  morekeywords=[1]{
    lp.setup, mv, lw, p.lw, sw, p.sw, pv.sdotsp.b, pv.shuffle2.b, p.subNR, p.addNR
  },
  morekeywords=[2]{
    zero, ra, sp, gp, tp, t0, t1, t2, t3, t4, t5, t6, s0, s1, a0, a1, a2, a3, a4, a5, a6, a7, a8, a9, a10, a11,
  },
  morestring=[b]",
  morestring=[b]',
}[strings, comments, keywords]

\lstdefinestyle{RISCV_STYLE}{
  language=RISCV,
  numbers=none,
  basicstyle=\scriptsize\ttfamily\color{code_default},
  keywordstyle=[1]\color{matplotlib0},
  keywordstyle=[2]\color{matplotlib1},
  float,
  captionpos=b,
  belowskip=-0.5cm
}

\usepackage{url}

\newcommand{\checkbox}{$\square$}%

\newcommand{\luca}[1]{{\textcolor{black}{#1}}}
\newcommand{\lu}[1]{{\textcolor{black}{#1}}}

\newcommand{\review}[1]{{\textcolor{black}{#1}}}
\newcommand{\reviewnew}[1]{{\textcolor{black}{#1}}}

\def\BibTeX{{\rm B\kern-.05em{\sc i\kern-.025em b}\kern-.08em
    T\kern-.1667em\lower.7ex\hbox{E}\kern-.125emX}}
\makeatletter
\def\ps@IEEEtitlepagestyle{%
  \def\@oddhead{}
  \def\@evenhead{}%
  \def\@oddfoot{%
    \vbox to0pt{\vss
      \hbox to\textwidth{%
        \parbox[t]{\textwidth}{\centering\scriptsize
          \copyright 2025 IEEE.  Personal use of this material is permitted.  Permission from IEEE must be obtained for all other uses, in any current or future media, including reprinting/republishing this material for advertising or promotional purposes, creating new collective works, for resale or redistribution to servers or lists, or reuse of any copyrighted component of this work in other works.
        }%
      }%
    }%
  }%
  \def\@evenfoot{}%
}
\makeatother
\begin{document}
\acrodef{AW}{alpha waves}
\acrodef{PULP}{Parallel Ultra Low Power}
\acrodef{SoC}{System on Chip}
\acrodef{BLE}{Bluetooth Low Energy}
\acrodef{EEG}{electroencephalography}
\acrodef{EMG}{electromyography}
\acrodef{ECG}{electrocardiogram}
\acrodef{PPG}{photoplethysmogram}
\acrodef{EOG}{electrooculography}
\acrodef{MM}{Motor Movement}
\acrodef{SSVEP}{Steady State Visually Evoked Potential}
\acrodef{HMI}{human-machine interface}
\acrodef{PMIC}{Power Management Integrated Circuit}
\acrodef{IMU}{inertial measurement unit}
\acrodef{DSP}{digital signal processing}
\acrodef{NN}{neural network}
\acrodef{ULP}{ultra-low-power}
\acrodef{sEMG}{surface electromyography}
\acrodef{ML}{Machine Learning}
\acrodef{LOSO CV}{leave-one-subject-out cross-validation}
\acrodef{CV}{cross-validation}
\acrodef{BCI}{Brain-Computer Interface}
\acrodef{CCA}{canonical-correlation analysis}
\acrodef{SoA}{State-of-the-Art}
\acrodef{SNR}{signal-to-noise ratio}
\acrodef{PGA}{programmable-gain amplifier}
\acrodef{BTE}{Behind-the-Ear}
\acrodef{CNN}{convolutional neural network}
\acrodef{ITR}{information transfer rate}
\acrodef{NE16}{Neural Engine 16}
\acrodef{PTH}{plated through hole}
\acrodef{CMRR}{common-mode rejection ratio}
\acrodef{AFE}{analog frontend}
\acrodef{PCB}{printed circuit board}
\acrodef{IFCN}{International Federation of Clinical Neurophysiology}
\acrodef{NCCA}{Normalized Canonical Correlation Analysis}
\acrodef{ADC}{Analog to Digital Converter}
\acrodef{CV}{Cross Validation}
\acrodef{SoA}{State-of-the-Art}
\acrodef{SIMO}{Single-Inductor Multiple-Output}
\acrodef{LDO}{Low Drop-Out}
\acrodef{PSRAM}{Pseudo Static Random Access Memory}
\acrodef{WLCSP}{Wafer Level Chip Scale Package}
\acrodef{PDM}{Pulse Density Modulation}
\acrodef{MLC}{Machine Learning Core}
\acrodef{AAD}{Acoustic Activity Detect}
\acrodef{ASC}{Adaptive Self-Configuration}
\acrodef{RNN}{Recurrent Neural Network}
\acrodef{SFU}{Smart Filtering Unit}
\acrodef{SDK}{Software Development Kit}
\acrodef{MCU}{Micro-Controller Unit}
\acrodef{SoC}{System on Chip}
\acrodef{JTAG}{Joint Test Action Group}
\acrodef{GPIO}{General Purpose Input-Output}
\acrodef{RMS}{Root Mean Square}
\acrodef{PTT}{Pulse Transit Time}
\acrodef{PAT}{Pulse Arrival Time}
\acrodef{IOT}{Internet of the Things}
\acrodef{AI}{Airtificial Inteligence}
\acrodef{BPM}{Blood Pressure Monitoring}

\acrodef{SSVEP}{Steady-state Visual Evoked Potential}
\acrodef{ASSR}{Auditory Steady-State Response}
\acrodef{SoC}{System-on-Chip}
\acrodef{Resp}{Respiration}
\acrodef{Temp}{Temperature}
\acrodef{DD}{Drowsiness Detection}
\acrodef{FMG}{Force Myography}

\acrodef{OFA}{Once-For-All}
\acrodef{SIMD}{Single Instruction, Multiple Data}
\acrodef{ELU}{Exponential Linear Unit}
\acrodef{ReLU}{Rectified Linear Unit}
\acrodef{RPR}{Random Partition Relaxation}
\acrodef{MAC}{Multiply Accumulate}
\acrodef{DMA}{Direct Memory Access}
\acrodef{BMI}{Brain--Machine Interface}
\acrodef{BCI}{Brain--Computer Interface}
\acrodef{SMR}{Sensory Motor Rythms}
\acrodef{EEG}{Electroencephalography}
\acrodef{SVM}{Support Vector Machine}
\acrodef{SVD}{Singular Value Decomposition}
\acrodef{EVD}{Eigendecomposition}
\acrodef{IIR}{Infinite Impulse Response}
\acrodef{FIR}{Finite Impulse Response}
\acrodef{FC}{Fabric Controller}
\acrodef{NN}{Neural Network}
\acrodef{MRC}{Multiscale Riemannian Classifier}
\acrodef{FLOP}{Floating Point Operation}
\acrodef{SOS}{Second-Order Section}
\acrodef{IPC}{Instructions per Cycle}
\acrodef{TCDM}{Tightly Coupled Data Memory}
\acrodef{FPU}{Floating Point Unit}
\acrodef{FMA}{Fused Multiply Add}
\acrodef{ALU}{Arithmetic Logic Unit}
\acrodef{DSP}{Digital Signal Processing}
\acrodef{GPU}{Graphics Processing Unit}
\acrodef{SoC}{System-on-Chip}
\acrodef{MI}{Motor-Imagery}
\acrodef{CSP}{Commmon Spatial Patterns}
\acrodef{PULP}{parallel ultra-low power}
\acrodef{SoA}{state-of-the-art}
\acrodef{BN}{Batch Normalization}
\acrodef{ISA}{Instruction Set Architecture}
\acrodef{ECG}{Electrocardiogram}
\acrodef{RNN}{recurrent neural network}
\acrodef{CNN}{convolutional neural network}
\acrodef{TCN}{temporal convolutional network}
\acrodef{EMU}{epilepsy monitoring unit}
\acrodef{ML}{Machine Learning}
\acrodef{DL}{Deep Learning}
\acrodef{AI}{Artificial Intelligence}
\acrodef{TCP}{Temporal Central Parasagittal}
\acrodef{LOOCV}{Leave-One-Out Cross-Validation}
\acrodef{WFCV}{Walk-Forward Cross-Validation}
\acrodef{RWCV}{Rolling Window Cross-Validation}
\acrodef{IoT}{Internet of Things}
\acrodef{AUC}{Area Under the Receiver Operator Characteristic}
\acrodef{DWT}{Discrete Wavelet Transform}
\acrodef{FFT}{Fast Fourier Transform}
\acrodef{TPOT}{Tree-based Pipeline Optimization Tool}
\acrodef{TUAR}{Temple University Artifact Corpus}
\acrodef{TUEV}{Temple University Event Corpus}
\acrodef{AEP}{Auditory Evoked Potential}

\acrodef{BSS}{Blind Source Separation}
\acrodef{ICA}{Independent Component Analysis}
\acrodef{ICs}{Independent Components}
\acrodef{ASR}{Artifact Subspace Reconstruction}
\acrodef{PCA}{Principal Component Analysis}
\acrodef{GAP}{Global Average Pooling}
\acrodef{FCN}{Fully Connected Networks}
\acrodef{MLP}{Multi-Layer Perceptron}
\acrodef{NAS}{Neural Architectural Search}
\acrodef{FP/h}{False Positives per Hour}
\acrodef{BVP}{Blood volume Pulse}
\acrodef{EDA}{Electrodermal Activity}
\acrodef{ACC}{Accelerometer}
\acrodef{CAE}{Convolutional Autoencoder}
\acrodef{SSWCE}{Sensitivity-Specificity Weighted Cross-Entropy}
\acrodef{CE}{Cross-Entropy}
\acrodef{PPG}{Photoplethysmography}
\acrodef{NIRS}{Near-infrared Spectroscopy}

\acrodef{SBP}{systolic Blood Pressure}
\acrodef{SVM}{Support Vector Machine}
\acrodef{LDA}{Linear Discriminant Analysis}
\acrodef{MSE}{Multiscale Entropy}
\acrodef{EMD}{Empirical Mode Decomposition}
\acrodef{HR}{Heart Rate}
\acrodef{GSR}{Galvanic Skin Response}
\acrodef{NFC}{Near Field Communication}
\acrodef{RR}{Respiration Rate}
\acrodef{RSP}{Respiration}
\acrodef{BCM}{Body Composition Measurement}
\acrodef{SKT}{Skin Temperature}
\acrodef{ICG}{Impedance Cardiography}
\acrodef{AFE}{Analog-Front-End}

\newacronym{spo2}{SpO\ensuremath{_2}}{Peripheral capillary oxygen saturation}


\title{BioGAP-Ultra: A Modular Edge-AI Platform for Wearable Multimodal Biosignal Acquisition and Processing}

\author{Sebastian~Frey, \IEEEmembership{Graduate Student Member, IEEE}, Giusy~Spacone, \IEEEmembership{Graduate Student Member, IEEE}, Andrea~Cossettini, \IEEEmembership{Senior Member, IEEE}, Marco~Guermandi, Philipp Schilk, Luca~Benini, \IEEEmembership{Fellow, IEEE}, and Victor~Kartsch, \IEEEmembership{Member, IEEE}
\thanks{This project was supported by the Swiss National Science Foundation (Project PEDESITE) under grant agreement 193813, by the ETH-Domain Joint Initiative program (project UrbanTwin), and by the ETH Future Computing Laboratory (EFCL). }
\thanks{Sebastian Frey, Victor Kartsch, Philipp Schilk, Luca Benini, and Andrea Cossettini are with the Integrated Systems Laboratory, ETH Z{\"u}rich, Z{\"u}rich, Switzerland.}
\thanks{Luca Benini is also with the DEI, University of Bologna, Bologna, Italy.}
\thanks{Marco Guermandi is with the DEI, University of Bologna, Bologna, Italy, and with EssilorLuxottica, Milano, Italy.}
}

\maketitle

\thispagestyle{IEEEtitlepagestyle}  

\begin{abstract}

The growing demand for continuous physiological monitoring and human-machine interaction in real-world settings calls for wearable platforms that are flexible, low-power, and capable of on-device intelligence. This work presents BioGAP\lu{-Ultra}, an advanced multimodal biosensing platform that supports synchronized acquisition of diverse electrophysiological and hemodynamic signals such as EEG, EMG, ECG, and PPG while enabling embedded AI processing at state-of-the-art energy efficiency. \mbox{BioGAP-Ultra} is a major extension of our previous \review{BioGAP} design aimed at meeting the rapidly growing requirements of wearable biosensing applications. It features (i) increased on-device storage ($\times$2 SRAM, $\times$4 FLASH), (ii) improved wireless connectivity (\review{supporting up to} 1.4 Mbit/s bandwidth, \luca{$\times$4 higher than BioGAP}), (iii)  enhanced number of signal modalities \luca{(from 3 to 5)} and analog input channels (\luca{$\times$2}).  Further, it is \review{accompanied} by a real-time visualization and analysis software suite \review{that supports} the hardware design, providing access to raw data and real-time configurability on a mobile phone. 
Finally, we demonstrate the system's versatility through integration into various wearable form factors: an EEG-PPG headband consuming 32.8 mW, an EMG sleeve at 26.7 mW, and an ECG-PPG chestband requiring only 9.3 mW \review{for continuous acquisition and streaming}, tailored for diverse biosignal applications. \review{To showcase its edge-AI capabilities, we further deploy two representative on-device applications: (1) ECG-PPG-based PAT estimation at 8.6 mW, and (2) EMG-ACC-based classification of reach-and-grasp motion phases, achieving 79.9 \% $\pm$ 5.7 \% accuracy at 23.6~mW.}
\luca{All hardware and software design files are also released open-source with a permissive license.}

\end{abstract}

\section{Introduction}


Recent advances in hardware miniaturization and embedded AI are driving the integration of biosignal monitoring into out-of-the-lab environments. Compact, low-power wearable devices now allow long-term recording of high-quality biosignals in daily life conditions, driving a global market projected to grow beyond USD 50 billion by 2030 \cite{GrandViewResearch_WearableMedicalDevices_2024}.
However, many existing platforms remain limited to single-sensor acquisition or offload processing externally, reducing robustness and missing out on opportunities for context-aware real-time interaction.


Multimodal biosignal acquisition has emerged as a key strategy to address these challenges. 
Combining complementary signals such as \ac{EEG} for cortical activity, \ac{EMG} for muscular function, \ac{ECG} for cardiac dynamics, and \review{\ac{PPG}} for peripheral hemodynamics improves the estimation of physiological states and mitigates the \lu{impact of motion artifacts and} noise \cite{uribe_2018_sensor_fusion}. 
However, the design of synchronized, wearable multimodal systems is still underexplored, especially in the context of multimodal \ac{BCI}s \cite{li2025multimodal}, where EEG poses additional challenges such as guaranteeing high signal quality with the low \ac{SNR}. This paper targets \lu{tasks requiring multimodal, multichannel biosignal acquisition (including EEG) and in-situ processing.}


On the algorithmic side, recent advances in \ac{ML} \lu{at} the edge have shown the feasibility of extracting meaningful biomarkers directly from raw data \cite{DaviesHarryJ.2024TDFR}, enabling real-time operation, privacy preservation, stand-alone operation, and enhanced battery life. Yet, transitioning these methods to wearable platforms featuring multiple signal sources and requiring computationally intensive processing remains challenging. 
%
Conventional \lu{\acp{MCU}} typically lack the computational performance and energy efficiency required to run real-time AI workloads under tight energy constraints, while dedicated \ac{SoC}s for AI acceleration often sacrifice flexibility or ease of integration \cite{song_769_2019, machetti_heepocrates_2024}.  
In contrast, \lu{multi-core \ac{MCU}} platforms such as GAP9 \cite{GAP_SDK} address this trade-off by combining general-purpose programmable RISC-V cores with a tightly coupled neural engine, offering both flexibility for general-purpose \lu{parallel} signal processing and efficient acceleration of \ac{ML} tasks within \lu{an integrated, heterogeneous} ultra-low-power \lu{multicore} architecture.

\begin{figure}[htb]
    \centering
    \includegraphics[width=1\columnwidth, trim={0cm 0cm 0cm 0cm},clip]{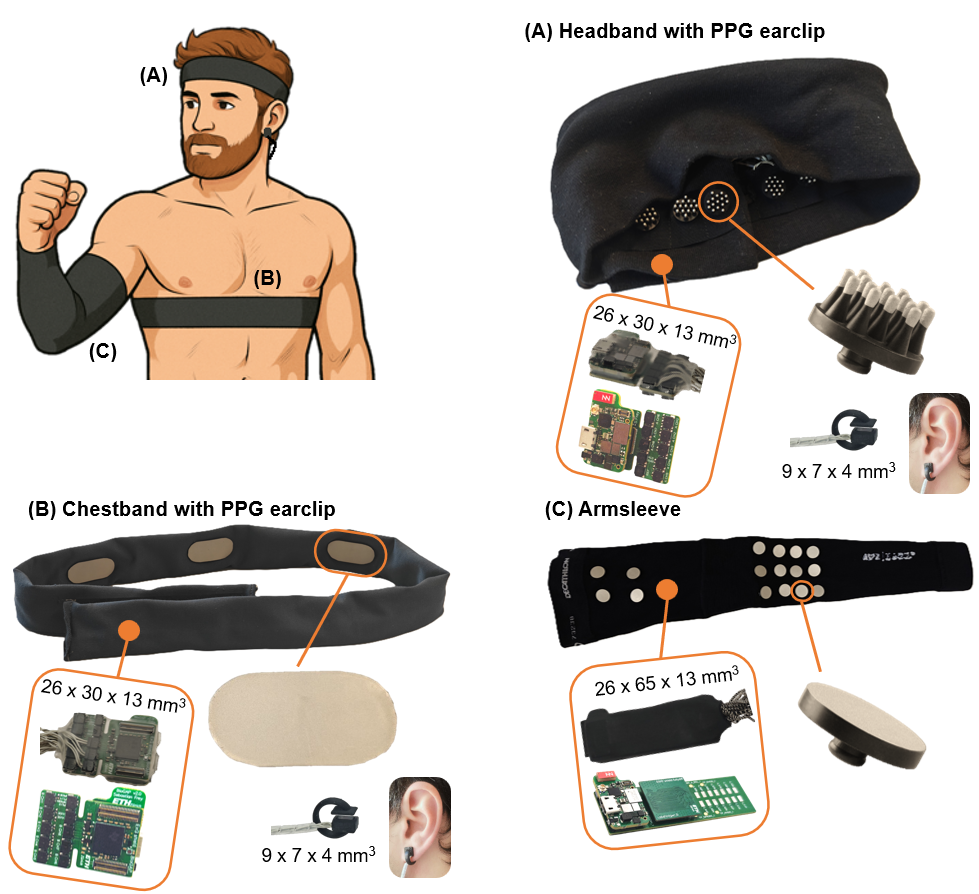}
    
    \caption{\luca{Overview of the proposed system and integration in multiple form factors. Top left: sketch of a person wearing the devices. (A) headband: the BioGAP\lu{-Ultra} platform is integrated in a textile headband with 16 SoftPulse dry EEG active electrodes, which can be coupled to a compact PPG sensing board on the earlobe; (B) chestband: the BioGAP\lu{-Ultra} platform is integrated in a chestband with flat, dry electrodes for ECG measurements; (C) arm sleeve: BioGAP\lu{-Ultra} uses the EMG sensing board and is integrated in a sleeve with flat and fully-dry EMG channels over the forearm (12 channels) and \reviewnew{upper arm} (4 channels).}}
    \label{fig:Overview}
\end{figure}

\lu{We aim at addressing the challenge of increasing multimodality and number of biosignal input channels, compounded with increased edge processing workloads. We introduce BioGAP-Ultra, which significantly extends BioGAP \cite{frey_2023_BioGAP}. BioGAP-Ultra is a scaled-up,}
compact, modular, and open-source platform for synchronized multimodal biosignal acquisition and on-device inference. \mbox{BioGAP\lu{-Ultra}} integrates flexible \ac{AFE}s with the GAP9 \ac{SoC} in a modular and configurable setup, and its versatility is demonstrated across different form factors (headbands, chestbands, \review{sleeves}) and use-cases (\luca{see Fig.~\ref{fig:Overview}}). 
%
We propose the following contributions:

\begin{itemize}
    
    \item Design of a compact and expandable computational node based on the GAP9 \ac{SoC} for AI-edge processing at \ac{SoA} energy efficiency across diverse heterogeneous applications, offering up to 15.6 GOPs for DSP and 32.2 GMACs for machine learning at 370 MHz clock frequency. \luca{With respect to BioGAP v1 \cite{frey_2023_BioGAP}, the new baseboard offers expanded RAM (256 to 512 Mbit) and flash (128 to 512 Mbit) capacity for larger edge-ML models, employs the nRF5340 for enhanced wireless throughput, and can be connected to a larger number of sensor front ends.}
    \item Design of multimodal biosignal acquisition modules for \ac{EEG}, \ac{EMG},  \ac{ECG}, and \review{\ac{PPG}}, as well as standard hardware (PCB) templates to expand other applications and use cases. \luca{Compared to \cite{frey_2023_BioGAP}, BioGAP-Ultra doubles the number of channels for ExG measurements} \lu{and raises modalities from 3 to 5 when including base board sensors \luca{(thus enabling 16-ch. ExG, PPG, accelerometry, Qvar, audio)}.}
    \item Design of a complete software suite for data visualization, storage, and signal analysis for mobile devices.
    \item Open-source release with a permissive license of all hardware and software design files, enabling the community to reproduce and build upon the design \href{https://github.com/pulp-bio/BioGAP}{https://github.com/pulp-bio/BioGAP}.
    \item System integration into three different form factors for fully-wearable synchronized multimodal data acquisition: an \ac{EEG} - \ac{PPG} headband, a chestband for \ac{ECG} and \ac{PPG} acquisition, and an \ac{EMG} sleeve for combined monitoring of the movement of the forearm and upper arm via EMG and 3D-accelerometry.
    \item Demonstration of biosignal acquisition at ultra-low-power across multiple wearable form factors: EEG-PPG headband (32.8 mW \review{in streaming mode}), EMG sleeve (26.7 mW \review{in streaming mode}), and ECG-PPG chestband (9.3 mW \review{in streaming mode}). \review{Additionally, we demonstrate edge deployment and characterization of two applications: ECG-PPG chestband-based \ac{PAT} estimation (8.6 mW) and EMG+ACC sleeve-based classification of the different phases of a continuous reach-and-grasp arm movement, achieving an accuracy of 79.9\% $\pm$ 5.7\% (23.6 mW)}.
 \end{itemize}

\section{Related Works}\label{sec:related_work}
\begin{figure}[htb]
    \centering
    \includegraphics[width=1\columnwidth, trim={0cm 0cm 0cm 0cm},clip]{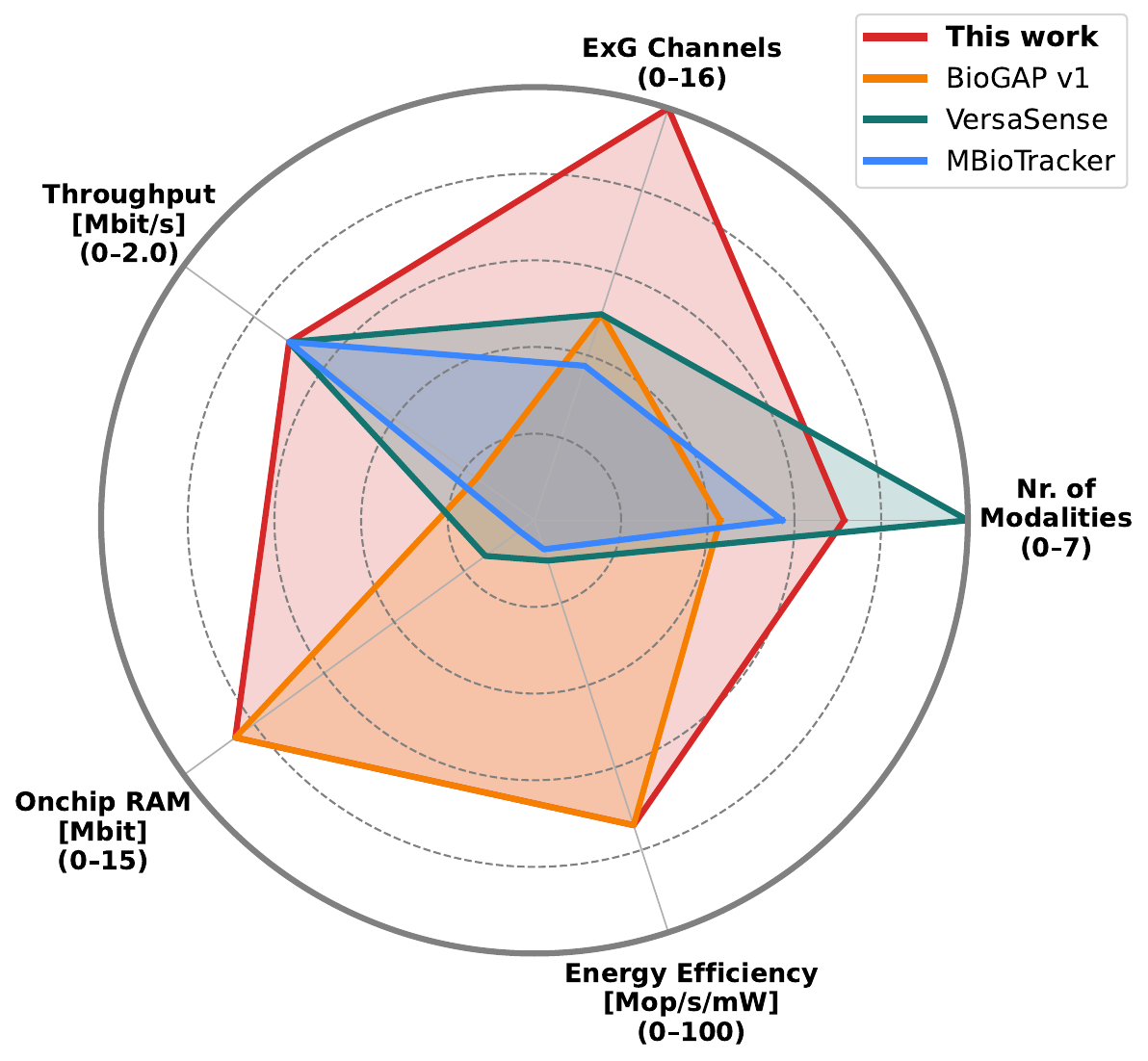}
    
    \caption{\luca{Normalized radar chart \review{quantitatively} comparing BioGAP\lu{-Ultra} (this work) to BioGAP v1 \cite{frey_2023_BioGAP} and to existing \ac{SoA} platforms across five main metrics (throughput, channel count, number of modalities, onboard memory, and onboard computing efficiency). Considering the total area covered, BioGAP\lu{-Ultra} outperforms all \ac{SoA} systems.}}
    \label{fig:radar_chart}
\end{figure}

Demonstrations of the advantages of multimodality span multiple domains, including \ac{BCI}s \cite{shatilov_emerging_2021,von_luhmann_m3ba_2017, almajidy_case_2023}, epilepsy monitoring \cite{ingolfsson2024brainfusenet, onorati2021prospective,zhang2024multimodal}, drowsiness detection \cite{frey_wearable_2024,mai2025multimodal,cao2025optimized}, neuromuscular monitoring \cite{9758822,zhang2024hand,kim2025development}, and blood pressure estimation \cite{sun2023systolic,tian2024flexible,hsiao2025predicting}.
This section reviews previous works on multi-modal, wearable sensing platforms, focusing on their evolution over time, specifications, sensing, and edge processing capabilities. 
In particular, we emphasize the challenges for multi-modal platforms capable of recording low-SNR signals like EEG \cite{casson_wearable_2008}. Table \ref{table:soa_comparison} summarizes key prior works \review{and Fig.~\ref{fig:radar_chart} quantitatively compares the main metrics}.  
%
%
%
\begin{table*}[h!]
\begin{center}
\begin{threeparttable}[b]
\caption{\review{Qualitative description of} state-of-the-art multimodal wearable platforms}
\label{table:soa_comparison}
\scriptsize{
\begin{tabular}
{
>{\centering\arraybackslash}m{0.25in} 
>{\centering\arraybackslash}m{0.7in} 
>{\centering\arraybackslash}m{0.2in} 
>{\centering\arraybackslash}m{0.35in} 
>{\centering\arraybackslash}m{0.7in} 
>{\centering\arraybackslash}m{0.35in} 
>{\centering\arraybackslash}m{0.3in} 
>{\centering\arraybackslash}m{0.25in} 
>{\centering\arraybackslash}m{0.2in} 
>{\centering\arraybackslash}m{0.4in} 
>{\centering\arraybackslash}m{0.9in} 
}

\toprule[0.20em]
\textbf{Paper} & \textbf{Modalities}  & \textbf{Modular} & \textbf{Synchronization} & \textbf{Form Factor} & \textbf{SoCs} & \textbf{Connectivity} & \textbf{\review{Open-source}} & \textbf{AI-ready} & \textbf{Power\footnotemark[12]} & \textbf{\review{Applications\footnotemark[13]}} \\ 
\midrule

\textbf{\cite{von_luhmann_m3ba_2017}} & ExG (6ch), fNIRS, ACC & \cmark& Wired &\makecell{4.2$\times$4.2$\times$0.6 cm$^3$}  & Cortex-M4 & BT (SPP) & \review{\cmark} & \xmark & $<$360 mW &\review{ \ac{AEP} (offline), Qualitative Multimodal Recording}\\


\midrule

\textbf{\cite{song_769_2019}} & ECG, BioZ, PPG, ACC & \xmark & Internal& Patch, size n/\reviewnew{a}  & Cortex-M4f & BLE & \review{\xmark} & \xmark & $769~\mathrm{\mu W}$ &\review{ On-chip HR, RR and SpO$_2$ estimation}\\

\midrule



\textbf{\cite{rosa_nfc-powered_2019}} & ECG, pH, Temp, Audio & \xmark & \makecell{Internal\footnotemark[3]}  & \makecell{Flexible \\ PCB}\makecell{5.0$\times$3.0$\times$0.5 cm$^3$} & PIC12 LF1822 & NFC & \review{\xmark} & \xmark & $1.8~\mathrm{mW}$ & \review{HR and systolic-interval estimation, sweat, pH, and skin temperature (off-device)} \\

\midrule

\textbf{\cite{schilk_vitalpod_2022}} & (HR, RR, SpO$_2$)\footnotemark[4], Temp, \review{IMU, QVAR} & \xmark & n/s  & \makecell{Earbud}\makecell{1.3$\times$2.1$\times$0.95 cm$^3$} & Apollo4 & BLE & \review{\xmark} & \xmark & $20.6~\mathrm{mW}$ & \review{HR, RR, SpO$_2$ estimation from in-ear PPG (offline)}\\

\midrule

\textbf{\cite{dellagnola_mbiotracker_2021}} & ECG, BMP\footnotemark[5], RSP, Temp & \xmark & Timestamp \& Interpolation  & \makecell{Chest band\footnotemark[4]} \makecell{4.2$\times$4.2$\times$0.6 cm$^3$} & STM32 L151 & BLE\footnotemark[6] & \review{\xmark} & \xmark & $97.42~\mathrm{mW}$\footnotemark[6] & \review{Cognitive-workload detection}\\

\midrule

\review{\textbf{\cite{Li2025}}} & \review{ExG, PPG, BioZ (RSP)} & \review{\xmark} &  \review{Hardware} & \review{\makecell{Patch} \makecell{50$\times$35$\times$13 cm$^3$}} & \review{STM32 WB5MMG} & \review{BLE} & \review{\xmark} & \review{\xmark} & \review{$30.4~\mathrm{mW}$} & \review{Qualitative Multimodal Recording, HR estimation (offline)}\\

\midrule

\textbf{\cite{najafi_versasens_2024}} & ExG (8ch), EDA, RSP, PPG, Temp, \review{IMU}, Audio & \cmark & Internal  & \makecell{Patch,\\ headband,\\ chest garment} \makecell{3.0$\times$5.5$\times$2.5 cm$^3$} & nRF5340 + \review{HEEPocrates SoC} & \review{BLE 5.4} & \review{\cmark} & \cmark & $155.4~\mathrm{mW}$\footnotemark[9] & Cough monitoring, heartbeat \& seizure detection\footnotemark[10] \\

\midrule
\luca{\textbf{\cite{frey_2023_BioGAP}}} & \luca{ExG (8ch), PPG, ACC} & \luca{\cmark} & \luca{Hardware} & \luca{\makecell{2.2$\times$1.5$\times$1.0 cm$^3$}} & \luca{GAP9 + nRF52811} & \luca{BLE 5.4} & \review{\xmark} & \luca{\cmark} & \luca{$35~\mathrm{mW}$} & \luca{HMI} \\
\midrule

\multirow{3}{*}[-7.8ex]{\reviewnew{\textbf{\makecell{This\\ work}}}} 
  & \reviewnew{\makecell{EEG (16ch), PPG,\\ ACC, QVAR, Audio}} 
  & \multirow{3}{*}[-7.8ex]{\reviewnew{\cmark}} 
  & \multirow{3}{*}[-7.8ex]{\reviewnew{Hardware}} 
  & \reviewnew{\makecell{Headband\\ 2.6$\times$3.0$\times$1.3 cm$^3$}} 
  & \multirow{3}{*}[-7.8ex]{\reviewnew{\makecell{GAP9 +\\ nRF5340}}} 
  & \multirow{3}{*}[-7.8ex]{\reviewnew{BLE 5.4}} 
  & \multirow{3}{*}[-7.8ex]{\reviewnew{\cmark}} 
  & \multirow{3}{*}[-7.8ex]{\reviewnew{\cmark}} 
  & \reviewnew{32.8~mW}
  & \reviewnew{\makecell{Neuro monitoring,\\ HMI}} 
\\
\cmidrule[0.1pt](lr){2-2}\cmidrule[0.1pt](lr){5-5}\cmidrule[0.1pt](lr){10-11}

  & \reviewnew{\makecell{EMG (16ch), ACC,\\ QVAR, Audio}} 
  &  
  &  
  & \reviewnew{\makecell{Armband\\ 2.6$\times$6.5$\times$1.3 cm$^3$}} 
  &  
  &  
  &  
  &  
  & \reviewnew{26.7~mW}
  & \reviewnew{\makecell{Gesture recognition,\\ muscular monitoring}} 
\\
\cmidrule[0.1pt](lr){2-2}\cmidrule[0.1pt](lr){5-5}\cmidrule[0.1pt](lr){10-11}

  & \reviewnew{\makecell{ECG (1ch), PPG,\\ ACC, QVAR, Audio}} 
  &  
  &  
  & \reviewnew{\makecell{Chestband\\ 2.6$\times$3.0$\times$1.3 cm$^3$}} 
  &  
  &  
  &  
  &  
  & \reviewnew{9.3~mW}
  & \reviewnew{\makecell{Cardiac monitoring,\\ activity assessment}} 
\\
\bottomrule[0.20em]

\end{tabular}
}
\vspace{0.1cm}
\begin{tablenotes}
\footnotesize
\item[\textsuperscript{1}] Estimated from reported average current draw of 0.6\,mA at 3.3\,V.
\item[\textsuperscript{2}] Data is segmented, only portions are saved.
\item[\textsuperscript{3}] Synchronized snapshots: data from all sensors is retrieved within a $5~\mathrm{s}$ window and transmitted over NFC.
\item[\textsuperscript{4}] Vital signs (HR, RR, SpO$_2$) are extracted off-device after transmission of raw data via BLE.
\item[\textsuperscript{5}] Also allowing for \ac{ICG}, \ac{EDA} and \ac{SKT} acquisition through the same front-end.
\item[\textsuperscript{6}] Estimated.
\item[\textsuperscript{7}] Observed but not specified. Measurement leads are not within the chest band.
\item[\textsuperscript{8}] Not for data streaming (only results are sent). An SD card is used for data storage.
\item[\textsuperscript{9}]Full-system power consumption measured in storage-stream mode with all sensor modules active.
\item[\textsuperscript{10}]Only demonstrated on pre-existing public dataset, end-to-end demonstration still missing.
\item[\textsuperscript{11}]Measured across three form factors (chestband, armband, headband). GAP9 contribution varies with AI workload.
\item[\textsuperscript{12}] \review{Values are not fully comparable due to differences in modality sets, channel counts, and operating points across systems. For details regarding the power numbers of this work, see Section V.}
\item[\textsuperscript{13}] \review{Applications experimentally demonstrated in each study. Off-device or offline processing is explicitly indicated. Otherwise, processing is on-device.}
\end{tablenotes}

\end{threeparttable}
\end{center}
\vspace{-8mm}
\end{table*}

Early attempts for multimodal sensing include the work of Von L\"uhmann et al. \cite{von_luhmann_m3ba_2017}, with a compact system \lu{(4.2$\times$4.2$\times$0.6~cm$^3$)} based on an ARM Cortex-M4 microcontroller that allows simultaneous recording of 6-channel ExG signals, 4-channel fNIRS, and accelerometer data, with BLE connectivity. The platform was validated through EEG-based auditory evoked potential studies and multimodal measurements captured during various activities, such as standing, walking, turning, and eye state differentiation (open vs. closed conditions). 
 However, the system requires up to $360~\mathrm{mW}$, without embedding AI-ready hardware for efficient processing.

Subsequent works showed further improvements in developing multimodal sensing systems while significantly reducing the power demands. For instance, Song et. al \cite{song_769_2019} have presented a multi-mode low-power \ac{SoC} coupled with a disposable patch for \ac{ECG}, Bio Impedance, \ac{PPG}, and \ac{ACC} readouts, including accelerators for extracting signal features (such as \ac{HR}), with a power consumption of $\sim$2\,$\mathrm{mW}$. 

Multi-modal sensing has also been featured in different form factors. Authors in \cite{rosa_flexible_2019} have developed a battery-less \review{chest} patch for cardiac, hemodynamic, and Endocrine parameter monitoring over \ac{ECG}, microphone, pH, and temperature sensors. Besides a compact form factor (the chest patch system is integrated within a flexible \ac{PCB}), the authors have included an NFC coil data transmission and energy harvesting over \ac{NFC} RF signals, allowing to reduce the device size while keeping the screening time to within $5~\mathrm{s}$. However, while these systems offer a convenient form factor and are suitable for basic screening applications, they lack sufficient temporal resolution, are not designed for continuous monitoring, and do not support \textit{edge-AI} capabilities. Authors in \cite{schilk_vitalpod_2022} have also explored using earbuds for \ac{HR}, \ac{RR},  \gls{spo2}, body temperature and \ac{ACC}. These parameters are computed offline over \ac{PPG}, \ac{Temp}, and \ac{IMU} sensor measurements, collected and wirelessly transmitted with a low-power BLE-capable \ac{MCU}, allowing for more than one day of operation.

While these works demonstrate improved wearability, they still lack power optimizations for long-term continuous monitoring or do not offer \textit{edge-AI} capabilities. Furthermore, these solutions primarily targeted general physiological monitoring (ECG, PPG, Resp, EDA), and did not consider more challenging biosignals such as \ac{EEG} and \ac{EMG}. 

An early attempt to exploit the embedded processing paradigm over multi-modal sensing is presented in \cite{dellagnola_mbiotracker_2021}. The device known as MbioTracker allows extracting human cognitive load onboard through signal pre-processing and \ac{SVM} classification of \ac{ECG}, \ac{RSP}, \ac{PPG}, \ac{BCM}/\ac{ICG}/\ac{EDA} and \ac{SKT} signals, leading to a fully stand-alone and energy-efficient solution ($4.03\text{--}19.0~\mathrm{mA}$). Still, while the authors have demonstrated the feasibility of onboard computation, the limited computational capabilities of the platform restrict it to lightweight processing, rendering it unsuitable for the more complex signal processing typically required for \ac{EEG} and \ac{EMG} acquisition, which the system does not support either.

Rapa et al. introduced GAPWatch \cite{rapa_embedded_2024}, a platform including the GAP9 \cite{GAP_SDK} \ac{SoC}, an energy-efficient \ac{PULP} processor featuring an \ac{NN} accelerator, a 16-channel \ac{AFE} for EMG, 2-channel PPG, a dedicated ECG, and IMU. This smartwatch is one of the first to provide a suitable platform for efficiently executing \textit{edge-AI} tasks over various signals. Despite lacking the flexibility to allow capturing other signals of the body, the platform allows for both long-term monitoring (BLE streaming) and online processing over the deployment of a real-time heart rate monitoring pipeline and advanced embedded multimodal analysis.

\review{A recent advancement in multimodal biosignal acquisition is presented in \cite{Li2025}, where the authors propose a platform capable of ExG, PPG, and bioimpedance measurements across different body sites. However, the system lacks embedded computational resources for on-device ML processing and is not designed for fully wearable, form-factor-integrated operation.}
\review{Another} recent advancement is the VersaSens platform \cite{najafi_versasens_2024},  a  highly modular, reconfigurable, and AI-capable wearable system designed for a wide range of biomedical applications and biosignals (such as ExG, \ac{ECG}, \ac{EDA}, \ac{RSP}, and others). VersaSens integrates a dual-core BLE-enabled processor (nRF5340) \review{on its main board module}, a reconfigurable \review{HEEPocrates} co-processor for efficient AI inference \review{(HEEPO board module)}, and multiple plug-and-play sensor modules that can be worn in different locations (chest, arm, shoulder, neck), depending on the application. The \textit{edge-AI} was validated for several tasks, including cough frequency monitoring, heartbeat classification, and seizure detection (yet only evaluating processing algorithms via public datasets, without a complete end-to-end validation). Still, while the system addresses many of the current issues for multimodal sensing, \lu{its size (3.0$\times$5.5$\times$2.5 cm$^3$)} limits usage in many applications requiring minimal form factors, and requires more than $130~\mathrm{mW}$ of power when running all modules. 

The previously mentioned works underscore the current challenges for multi-modal sensing, which can be summarized as follows:

\begin{itemize}
    \item \textbf{Form Factor:} most platforms remain bulky, rigid, or require cumbersome mounting (such as large acquisition boards or large enclosures), reducing suitability for unobtrusive or mobile use in real-world settings.
    
    \item \textbf{Support for ExG Signals:} several systems do not support EMG/EEG  modalities, due to their stringent requirements in terms of sampling rates, channel count, analog front-end noise, and synchronization, limiting the ability to fully leverage multi-modal acquisition for improved system performance.
    
    \item \textbf{Embedded Intelligence:} while most recent platforms are capable of performing on-board signal preprocessing, only a few are suitable for efficient \textit{edge-AI} processing, limiting real-time responsiveness, increasing dependence on external devices, and finally reducing the battery life of devices.
    
    \item \textbf{Power-Performance Trade-offs:} energy-efficient designs often achieve low power by reducing sensing frequency or coverage, impairing temporal resolution, and rendering them unsuitable for continuous or high-fidelity monitoring.
    
    \item \textbf{Limited Modularity and Extensibility:} most solutions are designed around fixed sensor configurations, making it difficult to adapt to diverse applications or body locations without extensive redesign. \review{Furthermore, the absence of open-source hardware and software frameworks in most prior works restricts reproducibility and community-driven development, slowing progress toward interoperable and customizable multimodal sensing ecosystems.}

    \item \textbf{Synchronization and Fusion Challenges:} multimodal fusion is rarely performed onboard and is hindered by the lack of precise synchronization mechanisms across sensing modalities.
    
    \item \textbf{Validation Scope:} while some systems claim real-time capability, many are validated only in offline or partial pipeline evaluations, lacking proper end-to-end validation under operational conditions.
    
\end{itemize}

\luca{Extending our previous conference paper which introduced BioGAP} \cite{frey_2023_BioGAP}, \luca{here we present}  BioGAP\lu{-Ultra}, a compact, modular, and \textit{edge-AI}-capable platform for multi-modal biosignal acquisition and processing. The device allows for synchronous biosignal data acquisition, including \ac{EEG}, \ac{EMG}, \ac{ECG}, and \ac {PPG}, and also supports the collection of \ac{ACC}, \lu{QVAR}, and acoustic (microphone) signals. The system is highly modular, enabling new features to be added over new \ac{PCB}s. The system is built around GAP9, a \ac{PULP} \ac{SoC} for efficient processing of computationally heavy algorithms. \luca{The radar chart in Fig. \ref{fig:radar_chart} illustrates the improvements introduced by BioGAP\lu{-Ultra} over previous systems, including its predecessor BioGAP v1, showing superior performance across most dimensions. Owing to the system’s modularity, this performance envelope can be further extended to support additional sensing modalities and edge-AI applications. A more detailed discussion on the comparison to state-of-the-art is provided in Sect.~\ref{sect:soa_comparison}.} In this work, we showcase three form factors enabled by the platform, namely, a chestband (\ac{ECG} + \ac{PPG}), a 16-ch \ac{EMG} sleeve, and a 16-ch headband for neural activity recording (\ac{EEG}). The design is coupled with software for signal acquisition and visualization (Android) that, together with the hardware, is released open-source with a permissive license, aiming to boost accessibility among the scientific community.

\section{System Design}\label{sec:system_design}

\subsection{Hardware}
BioGAP\lu{-Ultra} is designed to be a modular and easily expandable platform, allowing for quick system tailoring to the specific application, both in terms of sensing capabilities and form factor, without compromising performance. The core of the platform is a \emph{mainboard} that provides power management, control of sensors and sensor data flow, wireless connectivity, and computational capabilities on board. This mainboard is then connected to one or more expansion boards, which allocate the set of sensors specific to the target application. In this paper, the mainboard is combined with three expansion boards for EEG, EMG, and PPG sensing. A debugging board is also provided to easily connect the mainboard to the probes for programming the embedded processors and to expose the signals of the different interfaces to monitor and debug communication between the processors and the sensors.
\subsubsection{Mainboard}
\begin{figure}[htb]
    \centering
    \includegraphics[width=0.95\columnwidth, trim={0cm 0cm 0cm 0cm},clip]{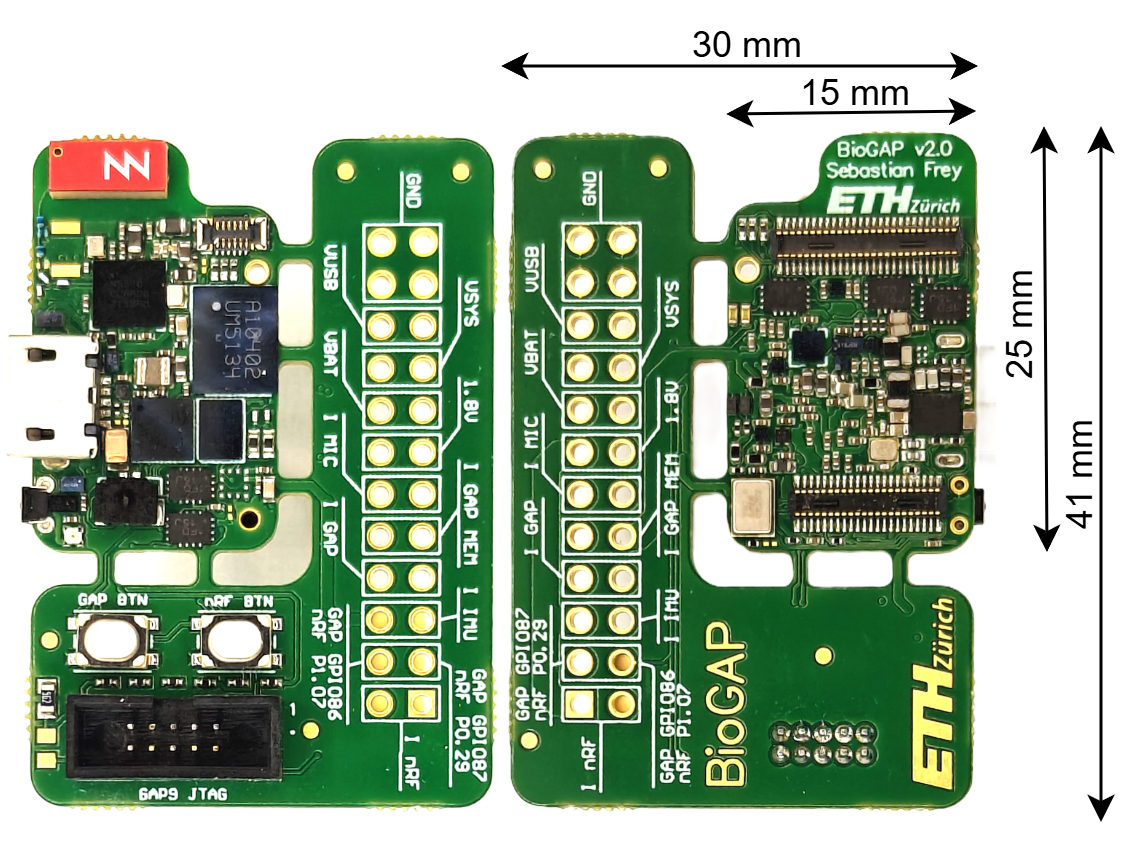}
    \caption{Photo of the Mainboard. The board comprises a breakaway section with pin headers and buttons to facilitate programming and debugging.}
    \label{fig:MB_PIC}
\end{figure}

\begin{figure}[htb]
    \centering
    \includegraphics[width=1\columnwidth, trim={0cm 0cm 0cm 0cm},clip]{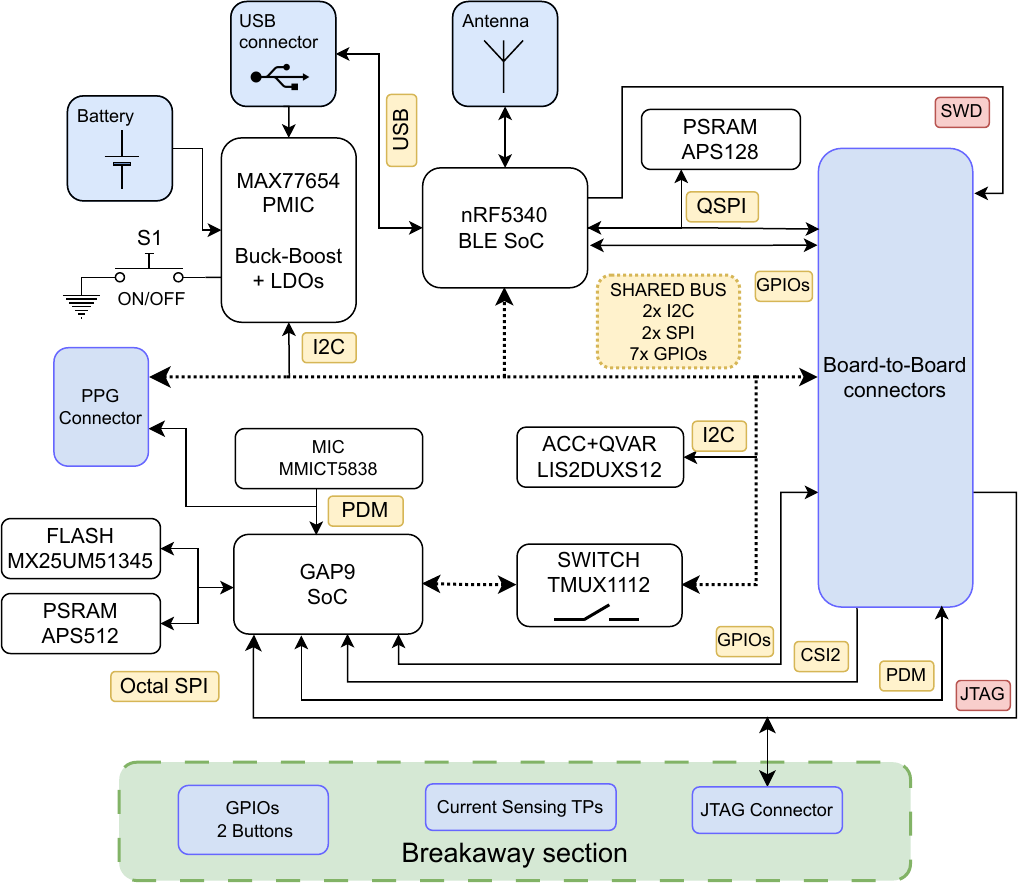}
    
    \caption{Block diagram of the Mainboard, with description of the interfaces. Power supply connections are not shown.}
    \label{fig:MB_BD}
\end{figure}

The mainboard \reviewnew{(see Fig. \ref{fig:MB_PIC} for a picture and Fig. \ref{fig:MB_BD} for a block diagram) has} a unipolar supply and common reference configuration, is the core of the platform\reviewnew{,} and is the only part that is always required in any configuration of the system. 
It comprises an ultra-low power \ac{PMIC}, MAX77654 from Analog Devices, which handles battery monitoring and charging, as well as generating five programmable voltages to supply the mainboard itself and the expansion boards. The \ac{SIMO} buck-boost architecture allows for generating three independent rails between 0.8 and 5.5 V in 50 mV steps, employing a single inductor to minimize area estate. Two \ac{LDO} regulators are also available with an output voltage between 0.8 and 3.975 V, programmable in 25 mV steps, providing ripple rejection for noise-sensitive applications, such as audio and biopotential acquisition. A bidirectional I2C interface allows for configuring and reading out the device's state. The battery is a compact  \SI{150}{mAh} LiPo battery, although it can be replaced by different batteries to balance battery life and size. A USB connector provides \lu{for an} external power source to recharge the battery.

Two \acp{SoC} are integrated, a Nordic nRF5340 for \ac{BLE} wireless connection and a GAP9 \ac{PULP} processor for energy-efficient \ac{NN} inferences and \ac{DSP}. As shown in Fig.~\ref{fig:MB_BD}, shared buses (2 I2Cs, 2 SPIs and 7 GPIOs) are used to control the system by the SOCs, which can both act as masters on the buses, and also to exchange data between the SOCs. Low-leakage switches (TMUX1112 from Texas Instruments) controlled from the nRF5340 can be used to isolate GAP9 IOs from the bus in case some of its IOs are switched off to save power, both in active and sleep states. 

The nRF5340 is a dual-core Bluetooth 5.4 \ac{SoC} supporting a wide range of wireless protocols, from Bluetooth LE and mesh to NFC, Thread, and Zigbee. \review{The nRF5340 was chosen for wireless connectivity due to its dual-core design, which allows concurrent sensor management and BLE 5.4 communication with high throughput.} The network processor (Cortex-M33) is clocked at 64 MHz and is optimized for low power and efficiency (101 CoreMark/mA), while the application processor (also Cortex-M33) is optimized for performance and adopts voltage-frequency scaling to be clocked at either 64 or 128 MHz. The maximum power consumption of the radio is 15.3 mW in TX mode at +3dBm output power and 9.3 mW in RX mode at 2 Mbps. \lu{Compared to BioGAP \cite{frey_2023_BioGAP}, this yields a 29\% improvement in wireless transmission efficiency.} The antenna is a NN03-320 from Ignion, ideal for space-constrained applications given its small form factor of (7 mm x 3 mm x 2 mm) and the lack of clearance requirement beyond the antenna footprint.
\reviewnew{The integrated 2.4~GHz transceiver supports BLE PHYs at 1~Mbps and 2~Mbps, as well as the LE Coded long-range PHY (500 and 125~kbps), allowing a configurable trade-off between throughput and link robustness. In addition, the \ac{SoC} provides a single-ended RF pin with an on-chip balun, so that only a compact external matching network is required to interface the NN03-320 antenna, which simplifies the RF layout and minimizes the occupied area.}
\review{Unless otherwise stated, in the following we operate the BLE using the 2M PHY (for improved energy efficiency over the 1M) and a connection interval of 7.5~ms.}

For applications requiring computational intensive tasks (such as \ac{ML} or advanced \ac{DSP}, \review{see the later Sect.~\ref{sect:emg_use_case}}), the full signal acquisition and processing is handled directly by the second \ac{SoC}, GAP9 \ac{PULP} processor, which combines a low power \ac{MCU} \lu{with} a programmable 9-core cluster and a hardware \ac{NN} accelerator. \review{We selected GAP9 for its ultra-low-power parallel architecture and hardware support for neural network acceleration, as well as efficient digital signal processing capabilities.} GAP9 can deliver up to 32.2 GMACs within a very low power envelope at 330 $\mu$W/GOP, thanks to adjustable dynamic frequency and voltage scaling and automatic clock gating. GAP9 integrates a highly flexible hardware accelerator, designed explicitly for handling \ac{NN} operations and basic kernels, optimized for inference of, e.g., \ac{CNN} and \ac{RNN}, and parallelized vector/matrix multiplications with asymmetric, scaled quantization. The GAP9 \ac{SDK} \cite{GAP_SDK} is equipped with a comprehensive kernel library and tools for quickly converting \ac{NN} and \ac{DSP} graphs into optimized code for neural networks and signal processing algorithms. 

An APS128040 \ac{PSRAM} from APmemory is available to the nRF5340 on a dedicated QSPI bus, providing 128 MBit of volatile data storage with transfer rates up to 576 MBit/sec for optional data buffering in case of a temporary \ac{BLE} throughput drop while streaming data. Since GAP9 can perform energy-efficient \ac{NN} inference on relatively large networks, a high-speed (3.2 Gbit/sec) and high-capacity (512 MBit) non-volatile flash memory (MX25UM51345 from Macronix) for code and weight storage is directly connected to GAP9 on Octal-SPI interface. An APS512XXN-OB9 \ac{PSRAM} (3.2 Gbit/sec, 512 MBit) from APmemory is connected to the same bus for volatile data storage and intermediate computation results. All three memories allow sleep states, which reduces the power consumption when not in use, and come in compact \ac{WLCSP} versions (sizes of 1.7~x~1.8~mm, 5.5~x~6.6~mm, 2.6~x~3.6~mm for APS128040, MX25UM51345, and APS512XXN-OB9, respectively).

The mainboard also includes an \review{accelerometer} and a microphone, which can be used for several applications. The LIS2DUXS12 is a 3-axis digital accelerometer with QVAR combining a very low supply current (2.7 $\mu$A in ultralow-power mode and 10.8 $\mu$A in high-performance mode) with features such as \ac{MLC} with \ac{ASC}. The dedicated internal engine can detect motions that can be used to control BioGAP\lu{-Ultra} (including free-fall, wake-up, tap recognition, and 6D/4D orientation) or to reject movement-related artifacts in the acquired sensor signals.
The microphone is a T5838 from TDK, featuring a digital \ac{PDM} output, ultralow-power operation, and \ac{AAD} with a current consumption as low as 20 $\mu$A, ideal for wake-words and voice control.

Two high-density low-profile board-to-board connectors (DF40C-40DP-0.4V and DF40C-50DP-0.4V) with a total of 90 available positions allow connecting the expansion boards to the main board. Overall, the following peripherals are available:
\vskip 0.2cm
\begin{itemize}
    \item 2 x I2C slave ports (master can be both GAP9 or nRF5340).
    \item 2 x SPI slave ports (master can be both GAP9 or nRF5340).
    \item 1 x single lane MIPI CSI-2 sink port connected to GAP9.
    \item 1 x PDM audio interface connected to GAP9.
    \item 1 x QSPI slave port connected to nRF5340.
    \item 14 x GPIOs connected to nRF5340.
    \item 13 x GPIOs connected to GAP9.
\end{itemize}
\vskip 0.2cm

To facilitate programming and debugging, the mainboard is fabricated with a breakaway section (see Fig. \ref{fig:MB_PIC}, upper and right side of the PCB). It allocates two test buttons (one per \ac{SoC}), a 20-pin connector exposing test points for supplies, current sensing of the different power domains, and 5 \ac{GPIO}s. A 10-pin 1.27 mm pitch JTAG connector is also provided for easy programming of GAP9. \review{During bring-up, this section provides measurement and debug access. After validation, it is removed to reduce the board to 15~x~25~mm. Once removed, the mainboard remains fully programmable by docking to the external debug board (Sec.~\ref{debug_pcb}), which re-exposes the GAP9 debug/program pins.}

\begin{figure}[htb]
    \centering
    \includegraphics[width=1\columnwidth, trim={0cm 0cm 0cm 0cm},clip]{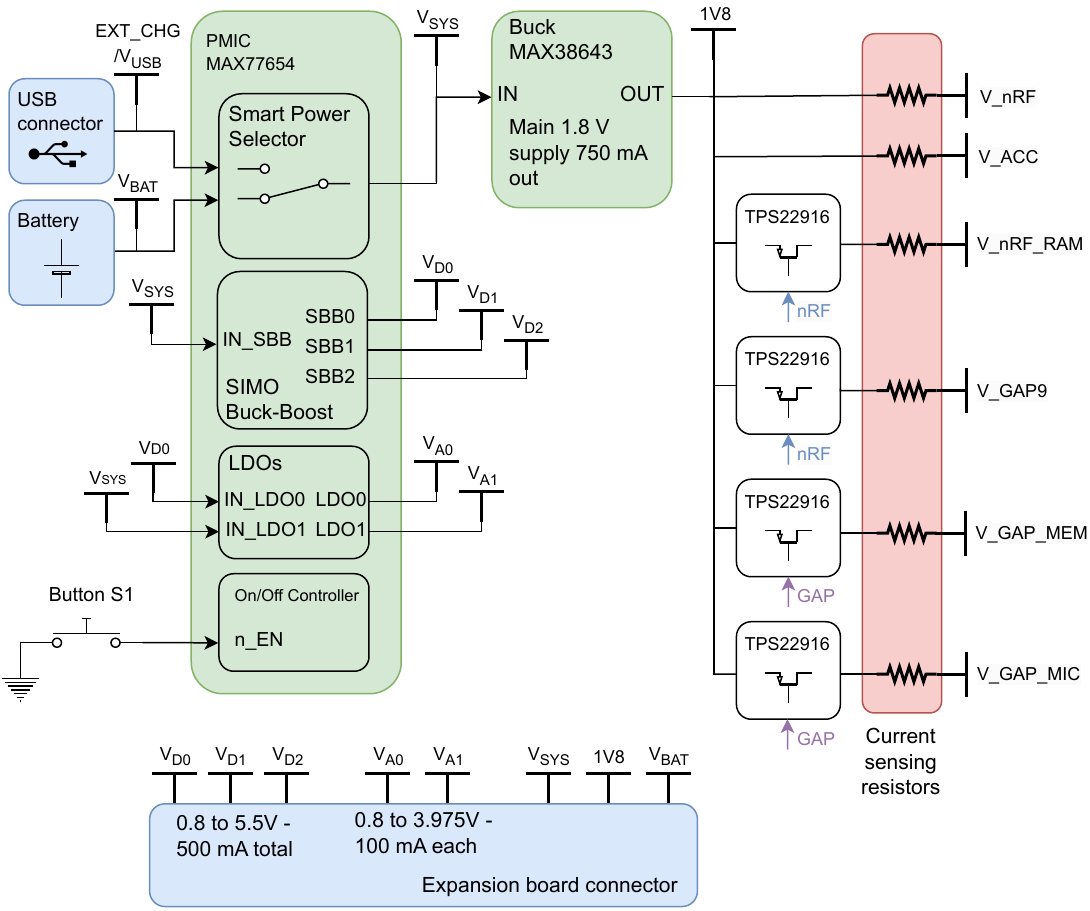}
    
    \caption{Block diagram of the Mainboard power management. The upper TPS22916 load switches are controlled by nRF5340 GPIOs, the lower ones by GAP9. V$_{SYS}$, V$_{D1}$, V$_{D2}$, V$_{D3}$, V$_{A1}$ and V$_{A2}$ are all controllable and programmable through a common I2C bus by both \ac{SoC}s.}
    \label{fig:MB_PM}
\end{figure}

The power management scheme of the platform is controlled by the mainboard and is depicted in Fig.\ref{fig:MB_PM}. The MAX77654 from Analog Devices provides integrated battery charging and power management for low-power applications. It features a \ac{SIMO} buck-boost regulator to provide three programmable power rails that require a single inductor to minimize area usage. Two \ac{LDO} regulators provide low-noise rails for audio and other noise-sensitive applications, such as biopotential acquisition. The capability to independently control the regulator outputs allows for selective power down of the rails depending on the platform's status, so as to minimize power consumption in every operating condition.  A bidirectional I2C serial interface allows for configuration and verification of the status of the PMIC from the SoCs, to selectively power on and off the different rails and configure their voltage levels. \review{An external button (S1 in Fig.~\ref {fig:MB_BD}) can be used to completely shut down the system by disconnecting the battery from the system through the PMIC Factory-Ship mode. The system can be re-enabled by pressing S1 again or by applying a valid USB voltage.}
The $V_{SYS}$ (2.7 to 5.5~V) output of the PMIC is connected to the input of a buck regulator (MAX38643 from Analog Devices), which provides a 1.8V rail to supply the mainboard. \review{This regulator is enabled whenever V$_{SYS}$ is present (i.e., a valid battery or USB voltage is applied, and the PMIC’s internal BATT–SYS path is closed).} Since the nRF5340 SoC and the \review{accelerometer} both present a negligible current in low-power and OFF modes (approximately 0.9 $\mu$A total current in OFF mode and 4 $\mu$A total current in low-power mode) and can be used as wake up source for the platform, these components are directly connected to the 1.8V supply (V$_{nRF}$ and V$_{\review{ACC}}$) through current sensing resistors. The other components on the mainboard can be switched ON and OFF through load switches (TPS22916 from Texas Instruments) which generate the following rails: V$_{nRF\_RAM}$ to supply the APS128 QSPI PSRAM, V$_{GAP9}$ for GAP9 SoC battery input and IO supply, V$_{GAP\_MEM}$ to supply the Octal-SPI Flash and PSRAM connected to GAP9, V$_{GAP\_MIC}$ to power the digital MIC. All of these rails have shunt current-sensing resistors to allow for power profiling of the main board. Every output rail of the PMIC (3 buck-boost regulator outputs, 2 LDO regulator outputs, V$_{SYS}$), V$_{BAT}$, and the 1.8V output of the MAX38643 are all made available to the expansion boards through the low-profile board-to-board connectors. \review{All components are integrated on an 8-layer PCB with a bare thickness of 1.6~mm and a maximum assembled height of 5.0~mm.}

\subsubsection{ExG sensing board}
\begin{figure}[htb]
    \centering
    \includegraphics[width=0.95\columnwidth]{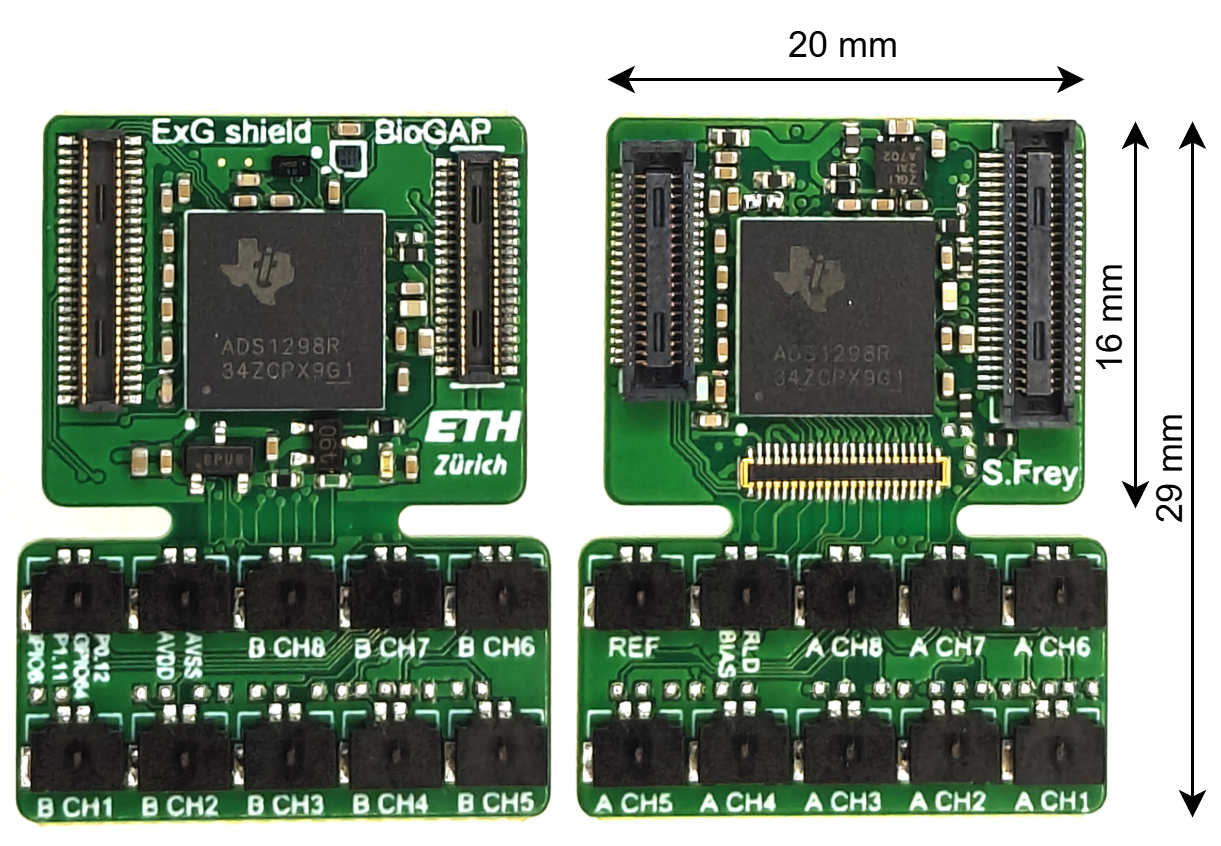}
    \caption{Top (left) and bottom (right) picture of the ExG PCB}
    \label{fig:EEGB_PIC}
\end{figure}

\begin{figure}[htb]
    \centering
    \includegraphics[width=1\columnwidth, trim={0cm 0cm 0cm 0cm},clip]{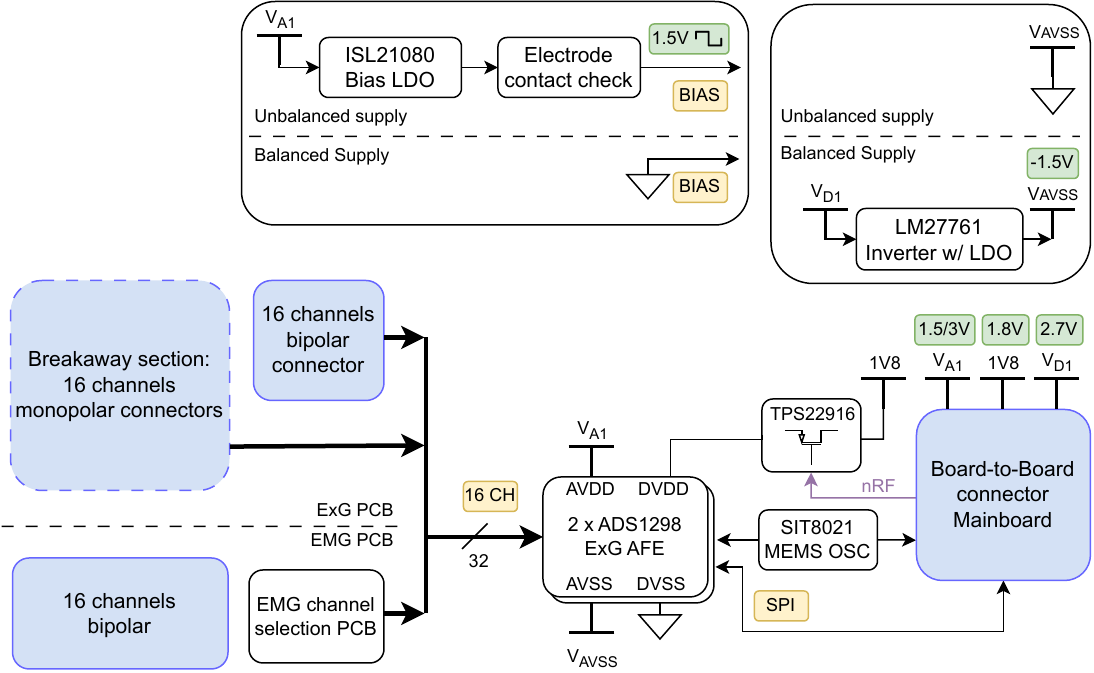}
    
    \caption{Block diagram of the ExG \review{(balanced or unbalanced supply)} \review{and EMG (balanced supply)} PCB.}
    \label{fig:BD_ExG_EMG_PCB}
\end{figure}

The ExG sensing board is depicted in Fig.~\ref{fig:EEGB_PIC}. At its bottom side, the board integrates the low-profile high-density connectors for mating with the main board, while an inverted replica is provided on the top side, with the same mechanical characteristics and pinout. This way, several expansion boards can be stacked to extend the system further \review{(note: the mechanical stacking limit is determined by the chosen connector height, which in the present design is 3 mm and can be increased up to 4 mm using higher-profile variants)}. These can be identical ExG sensing boards to extend the number of channels or boards with different types of sensors to extend the system's modalities.
To guarantee flexibility without compromising area, a high-density low-profile 42-position board-to-board connector (505070-4222 from Molex) is made available to connect the external electrodes with either bipolar or monopolar montage. At the same time, to improve functionality and ease of use, a breakaway section of the board allows for connecting directly the 16-channel electrodes (plus bias and reference) in monopolar configuration to single-channel connectors (DF52-2S-0.8H).
Two 8-channel 24-bit \ac{AFE} for bio-potential measurements (ADS1298 from Texas Instruments), perform ExG signal conditioning and analog-to-digital conversion. \review{We selected the ADS1298 for its excellent noise performance (4 {\textmu}Vpp) and CMRR (-115 dB)\footnote{\review{compatible with the low-noise requirements for EEG signals (see Sect.~\ref{sect:noise})}}, high sampling speed (up to 32 kSPS\footnote{\review{compatible with the $>500$ Hz bandwidth requirements of EMG signals}}), low power (0.75 mW/channel), programmable gain, and eight simultaneous acquisition channels, enabling scalable multi-signal recording.
} The AFEs can be configured at run time according to the specific application needs, with data rates ranging from 250 to 32 kSPS and a programmable gain of 1 to 12. Unused channels can be selectively turned off. \review{The two ADS1298s are connected in parallel on the SPI bus. Daisy-chain mode is not used. This allows operating only one ADS1298 when fewer channels are required, which helps reduce power consumption.}
\reviewnew{To ensure synchronized sampling across the two ADS1298 AFEs, a shared external clock is used. A dedicated 2.048~MHz oscillator (SIT8021AC-J3-18S-2.048000) provides the master clock to both devices. When additional ExG sensing boards are stacked, this clock is distributed through the board-to-board connector, ensuring that all ADS1298 devices operate synchronously in multi-board configurations.}
Mounting options allow for both single and dual-supply configurations (Fig.~\ref{fig:BD_ExG_EMG_PCB}). \review{Selection is made via different soldering configurations at build time (no runtime switching)}. Active and passive electrodes are supported for the single-supply configuration. Since the integrated lead-off detection of the ADS1298 can not be used with active electrodes, an external contact check circuitry \cite{Kartsch2020} is added to perform a contact quality check by injecting a small square wave on the bias electrode and detecting the presence and level of attenuation of such a wave in the acquired signals. \review{All components are integrated on a 6-layer PCB with a bare thickness of 1.0~mm and a maximum assembled height of 4.5~mm.}

\subsubsection{EMG sensing board}

\begin{figure}[htb]
    \centering
    \includegraphics[width=1\columnwidth, trim={0cm 0cm 0cm 0cm},clip]{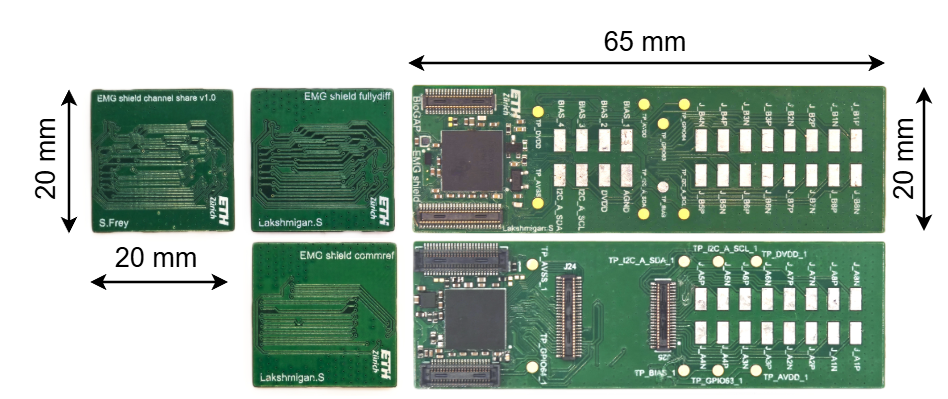}
    
    \caption{\review{Picture of the electrode selection boards (left) and the EMG sensing board (right).}}
    \label{fig:picture_EMG_pcb_sel_pcb}
\end{figure}

The EMG sensing board has an architecture similar to that of the ExG sensing board, making use of two ADS1298 \acp{AFE}, \review{ and also integrating all components in a 6-layer PCB with a bare thickness of 1.0~mm and a maximum assembled height of 4.5~mm.} \reviewnew{Similar to the ExG sensing board, the EMG board relies on the 2.048~MHz clock generated by the oscillator (SIT8021AC-J3-18S-2.048000) and distributed through the board-to-board connector. This ensures that all ADS1298 AFEs, whether on a single EMG module or across multiple stacked boards, operate synchronously.} It supports a dual-supply operation (see Fig.~\ref{fig:BD_ExG_EMG_PCB}), and employs identical board-to-board connectors, enabling flexible stacking of multiple sensing modules according to system requirements. A key differentiation of the EMG sensing board is its versatile electrode interfacing, designed for enhanced configurability, achieved through a dedicated electrode selection board that can be plugged into the EMG PCB. The modularity provided by this selection board facilitates quick adaptation to various electrode configurations, addressing diverse application demands effectively. \review{The selection board configures channels by shorting ADS1298 input pins. Electrode cables are soldered on the EMG PCB, and the selection board only reroutes inputs. Fig.~\ref{fig:picture_EMG_pcb_sel_pcb} shows a picture of the EMG sensing board alongside the electrode selection boards.}

The available electrode interfacing modes include:
\begin{itemize}
    \item Fully Differential: All electrodes are connected in fully differential mode to the respective input of the \ac{AFE}. This configuration is ideal in a setup where full flexibility is needed.
    \item Partial Electrode Share: Electrodes are selectively shared between pairs of channels. This configuration is ideal for scenarios where channel count and electrode count must be optimized due to physical constraints.
    \item Common Reference: All channels share a common reference electrode, reducing the number of electrodes and wiring. 

\end{itemize}
The selectable electrode configurations allow for easy system adjustment to the required setup.

\subsubsection{PPG sensing board}

\begin{figure}[htb]
    \centering
    \includegraphics[width=0.8\columnwidth, trim={0cm 0cm 0cm 0cm},clip]{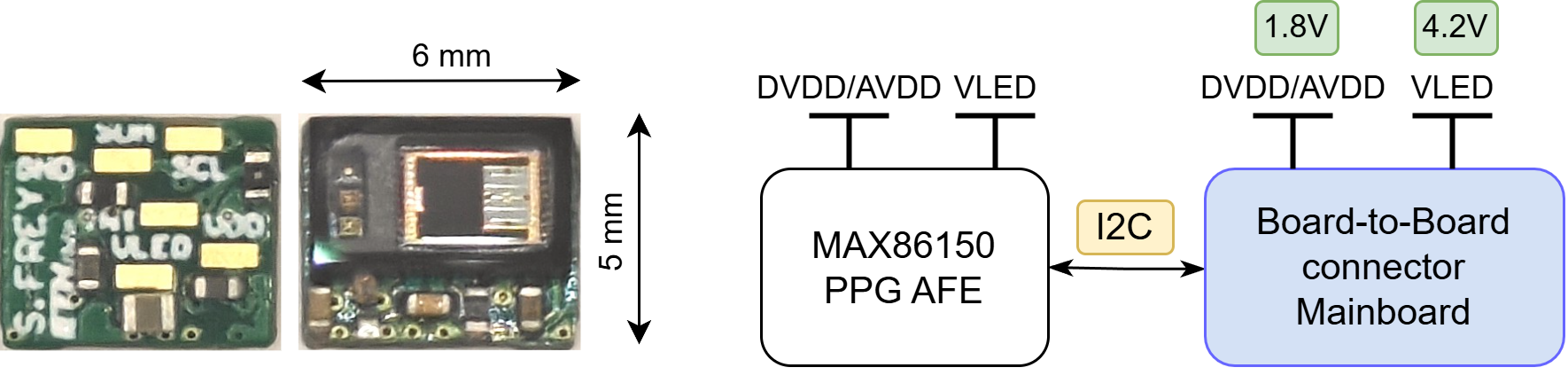}
    
    \caption{\review{Picture and} block diagram of the PPG PCB.}
    \label{fig:PPG_PCB}
\end{figure}

The PPG sensing board is based on the MAX86150 (Maxim, 3.3x5.6~mm$^2$). \review{We chose this component in light of the full integration of LEDs, photodetector, and low-noise electronics into a compact form-factor, which enables seamless integration into a compact earlobe clip.} The compact sensor and limited number of additional components allowed for a highly compact PCB design, measuring only 5~x~6~mm. The MAX86150 is a fully integrated bio-signal sensor capable of conducting PPG (pulse oximetry and heart rate sensor) measurements. It operates with a 1.8~V supply (digital and analog) and a 4.2~V LED supply voltage, drawing a current of 480~\textmu A~(typ.) at 10~SPS and \SI{1.15}{mA}~(typ.) at 100~SPS in SpO$_2$ mode. \review{The PPG sensor and passives are integrated on a 4-layer PCB with a bare thickness of 0.8~mm. The maximum assembled height is 2.8~mm, including the sensor. The board connects to the main board via a cabled connection that mates to a dedicated connector (see PPG connector in Fig.\ref{fig:MB_BD}), exposing the required power and digital I2C lines.}



\subsubsection{Debug board}
\label{debug_pcb}
The Debug PCB provides convenient access to all signals connecting the mainboard and sensing shields via pin headers. Both the mainboard and sensing shields can be directly connected to this debug PCB through their board-to-board connectors. Additionally, the debug PCB exposes flash and debug interfaces for both the nRF5340 and GAP SoCs, and includes three buttons, three status LEDs, and an RGB LED to facilitate debugging and device control. \review{All the components of this board are mounted on a 4-layer PCB, with a total height of 13~mm.}

\begin{figure}[htb]
    \centering
    \includegraphics[width=0.9\columnwidth, trim={0cm 0cm 0cm 0cm},clip]{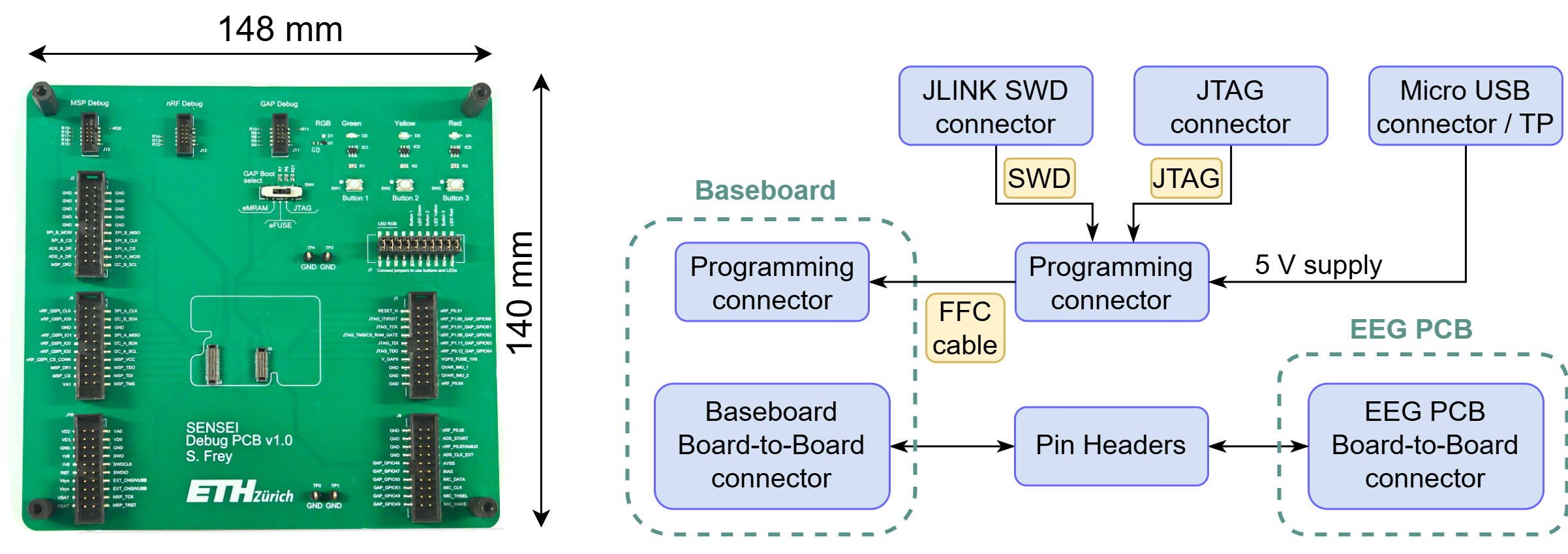}
    
    \caption{\review{Picture and block diagram of the Programmer PCB.}}
    \label{fig:PPG_PCB}
\end{figure}

\subsection{Firmware and Software Stack}\label{sec:FW_SW}

The platform can operate in two modes: Data streaming and Edge AI. In the data streaming mode, the nRF5340 directly transmits raw data to the receiving BLE device. In the edge-AI mode, GAP9 performs local processing, and the nRF5340 transmits only the processing outputs (e.g., classification outputs) at a reduced data rate.

\subsubsection{BioGAP-Ultra Firmware}

The firmware runs on the nRF5340 SoC using the Zephyr RTOS and follows a modular design. Each functionality is managed in a dedicated thread:

\begin{itemize}
    \item \textbf{State Machine Thread:} Manages the overall system state.
    \item \textbf{Power Management Thread:} Communicates with the MAX PMIC, controls power domains, monitors battery status, and provides information on charging state, voltages, and power.
    \item \textbf{BLE Advertising Thread:} Handles BLE advertising in peripheral mode.
    \item \textbf{BLE Send Thread:} Sends data over BLE when the message queue is not empty; otherwise, remains in sleep mode.
    \item \textbf{BLE Receive Thread:} Sleeps until a command is received (e.g., from a GUI), then executes the corresponding action.
    \item \textbf{\review{ACC} Thread:} Manages communication and data acquisition from the \review{accelerometer}.
    \item \textbf{PPG Thread:} Handles communication and data retrieval from the PPG sensor.
\end{itemize}

\subsubsection{BioGAP-Ultra Software}

BioGAP-Ultra \review{relies on} a software suite (Android-based) for data reception, visualization, and storage. 

\begin{figure}[htb]
    \centering
    \includegraphics[width=0.9\columnwidth, trim={0cm 0cm 0cm 0cm},clip]{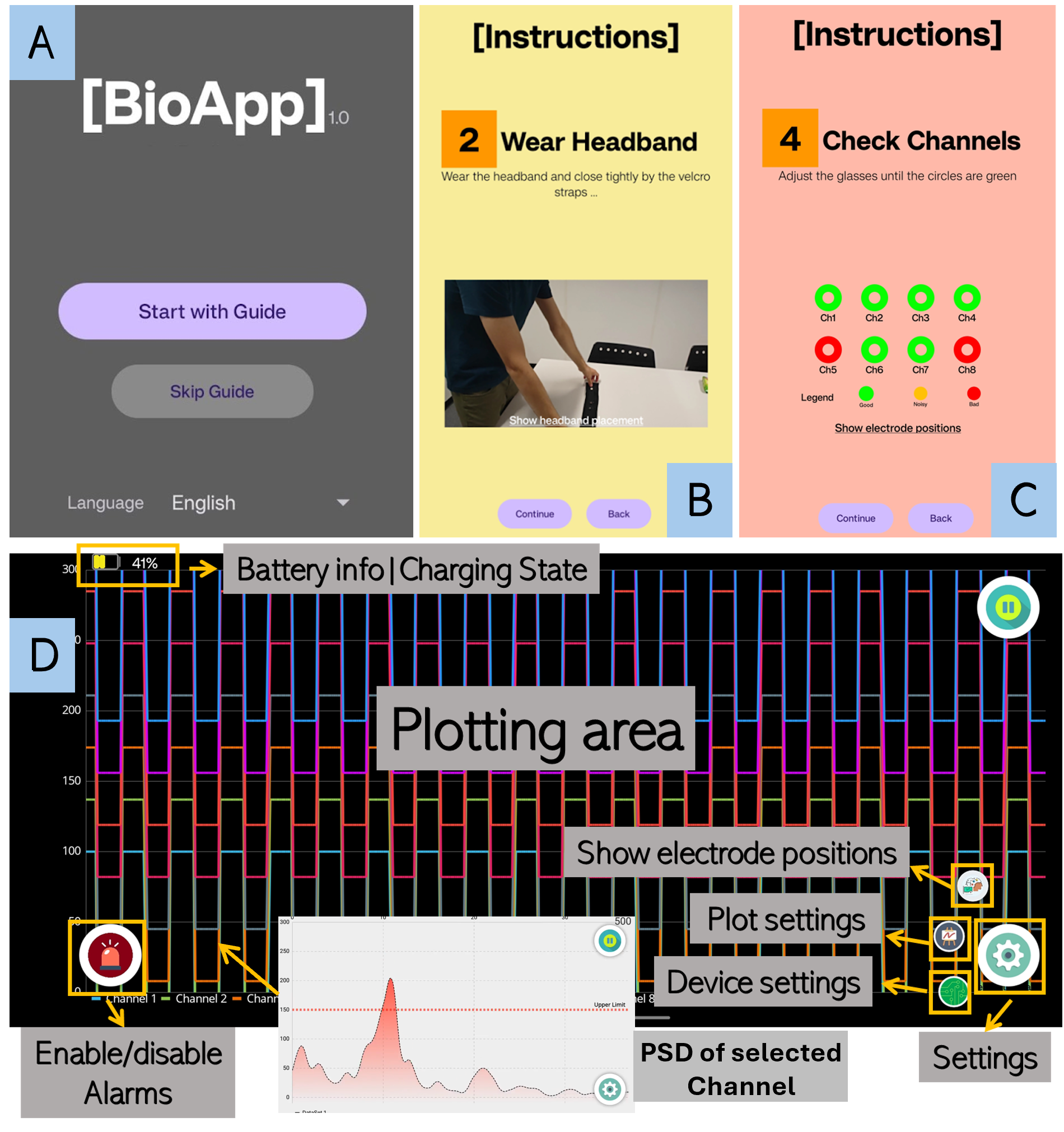}
    
    \caption{BioApp Android application. (A) shows the main screen, where users can choose to use or skip the usage tutorial while also allowing for selecting the language. (B) Shows one set of usage instructions. (C) shows the contact quality measurement results, while (D) shows the streaming visualization with all the controls available, including PSD visualization option. }
    \label{fig:Android_app}
\end{figure}

\subsection{Wearable Integration}\label{sec:WearableIntegration}

\subsubsection{EEG Headband \review{and PPG clip}}
\label{systemDesing:EEGHeadband}
The Mainboard with the ExG sensing board is fully integrated into a textile-based headband designed to be comfortable, flexible, and suitable for extended wear, as illustrated in Fig.~\ref{fig:Overview}. The ExG sensing board is configured to have a unipolar supply and common reference configuration, and it interfaces the EEG electrodes through the board-to-wire connectors on the break-away part of the PCB. The headband features 16 fully dry \ac{EEG} channels arranged in a common reference (monopolar) configuration. EEG measurements are conducted using dry pin electrodes (SoftPulse, Dätwyler Schweiz AG), evenly placed equidistantly around the headband to ensure consistent and balanced spatial coverage. \review{Following the 10 -– 20 system, electrodes are approximately at F7, FT7, T7, TP7, P7, PO7, O1, PO1, PO2, O2, PO8, P8, TP8, T8, FT8, and F8. This circumferential montage preserves a non-stigmatizing headband form and enables occipital, temporal, and partial frontal monitoring. Such a configuration has been shown suitable for applications including seizure detection using temporal EEG channels \cite{ingolfsson2024brainfusenet}, EEG-based speech imagery \cite{ingolfsson2025wearable}, and drowsiness detection \cite{frey2024wearable}. Reference and ground are placed frontally (similar to \cite{japaridze2023automated}), as the headband naturally keeps the dry electrodes in place. This can increase \ac{EOG} contamination and is an intentional wearability trade-off that can be mitigated through re-referencing or alternative electrode placement with wet electrodes on mastoids. 
}

To further enhance signal stability and minimize artifacts, the bias and reference electrodes are implemented as dry flat electrodes positioned on the forehead. The electrodes are active through a two-wire configuration\cite{kartsch_biowolf_2019}, which improves signal integrity by reducing susceptibility to external noise and motion artifacts while limiting the amount of cables needed.

The \ac{PPG} sensing board is \review{integrated into a 3D-printed clip and} placed on the earlobe, facilitating concurrent cardiovascular data acquisition. Additionally, BioGAP\lu{-Ultra} is embedded within the headband, providing a versatile, modular, and unobtrusive solution for real-world physiological monitoring applications, ensuring high-quality measurements in a user-friendly, comfortable form factor.

\subsubsection{EMG sleeve}
The EMG sensing board features fully differential electrode interfacing with a bipolar supply and is integrated into a fully flexible and stretchable textile sleeve. Dry flat electrodes (Dätwyler Schweiz AG) in passive configuration minimize cabling complexity and are interfaced using stretchable PhantomType X cables from Nanoleq \cite{nanoleq2024}. These stretchable cables enable the coverage of multiple body compartments, such as simultaneous recordings from the forearm and upper arm.

On the upper arm, differential electrode pairs are positioned over the long and lateral heads of the triceps and the long and short heads of the biceps, totaling four channels (8 electrodes). On the forearm, an array of electrodes (6 x 4 electrodes, 12 channels) is placed equidistantly around the upper and lower portions, covering the extensor and flexor muscle groups. 
\reviewnew{This redundant, closely spaced layout follows established principles of HD-EMG, which aim to uniformly cover the target muscle region with an approximately constant inter-electrode spacing. Such spatial distribution increases robustness to electrode placement variability, captures motor-unit action potential propagation across the muscle, and better represents 
heterogeneous activation patterns compared to single-site EMG recordings\cite{tacca_wearable_2024, tanzarella_arm_2024}.} The EMG sleeve facilitates monitoring muscle activity during functional movements commonly performed in daily tasks.

BioGAP is fully enclosed in the front part of the sleeve, enabling simultaneous tracking of orientation, acceleration, and overall limb dynamics during movement.

\subsubsection{ECG chestband}
The Mainboard with the ExG sensing board configured in the bipolar supply configuration is integrated into a stretchable textile chestband, designed for comfort and ease of wear. ECG measurements are enabled with a single-channel setup with passive dry sheet electrodes (Dätwyler Schweiz AG), oval-shaped with dimensions of 4x2 cm. The channel electrodes are placed on the right and left sides of the sternum, below the pectoral muscle line. The bias electrode is positioned on the left side, ensuring reliable skin contact. Additionally, the \ac{PPG} sensing board is attached to the earlobe, enabling simultaneous \ac{PPG} monitoring.

\subsubsection{Encapsulation and safety of form factors}
\review{All PCBs are conformally coated to prevent sweat/humidity ingress and corrosion. Modules are then encapsulated with heat-shrink tubing and integrated into textile pockets (headband, chestband, EMG sleeve) that provide mechanical protection and electrical isolation from the skin. The \ac{PPG} front end is housed in a dedicated clip with an optical window and light-blocking walls to maintain stable contact and reduce optical leakage. These measures increase safety, preventing user contact with conductive parts, improve comfort and durability, and enhance signal quality by limiting motion artifacts (ExG) and ambient-light interference (\ac{PPG}). The main bodies of all form factors are integrated into a stretchable fabric. See Fig.~\ref{fig:Overview} for the fully encapsulated assemblies.}

\section{System characterization}
\label{sect:results}

\subsection{Noise characterization}
\label{sect:noise}

We perform noise characterization of the ADS1298 operating at a sampling rate of 1~kSPS and a gain of 6 in high-resolution mode, with a \mbox{-3~dB} cutoff frequency of 262~Hz. The system exhibited an integrated \ac{RMS} noise of $0.47~\mu V$ in the frequency range of 0.5 to 100~Hz, which aligns with the standards set by the International Federation of Clinical Neurophysiology (IFCN) for recording \ac{EEG} signals in clinical settings \cite{nuwer1998ifcn}.

\subsection{Wireless characterization}

\begin{figure}[bt]
    \centering
    \includegraphics[width=1\columnwidth, trim={0cm 0cm 0cm 0cm},clip]{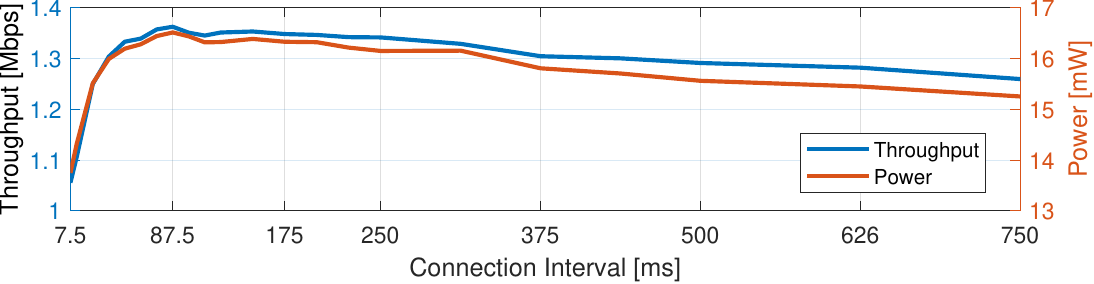}
    
    \caption{\review{Throughput/power for different connection intervals.}}
    \label{fig:wireless_characterization}
\end{figure}

\review{
We characterize the wireless connectivity by exploring different connection intervals, in the range of 7.5ms to 750ms. Using the 2M PHY with a connection interval of 87.5ms enables sustaining data rates as high as 1.362 Mbps (comparable with the SoC advertised throughput and online benchmarks \cite{Afaneh2023BLEThroughput}). The complete characterization is presented in Fig. \ref{fig:wireless_characterization}. The lowest connection interval of 7.5~ms guarantees a throughput of more than 1~Mbps while offering the lowest latency. This value will be used for all the following experiments.
}

\review{
When transmitting 16-ch. EMG and ACC data, the required data throughput is 216~kbps, which can be sustained by the lowest connection interval indicated above. With this datarate, we do not experience packet losses. 
}

\reviewnew{Under adverse BLE conditions, such as a noisy \SI{2.4}{GHz} environment or temporary body occlusion between transmitter and receiver, retransmissions of lost BLE packets are handled automatically by the BLE stack, thereby guaranteeing reliable data transmission at the application layer without the need to design custom error handling. Inevitably, retransmission increases radio-on time and power consumption, while reducing the maximum sustainable throughput. However, across all measurements reported in this work, the link quality remained high, and no data loss was observed.}

\begin{figure*}[h!]
    \centering
    \subfloat[][(a) NCCA response to target and non-target trials]{
        \includegraphics[width=0.9\textwidth,page=1]{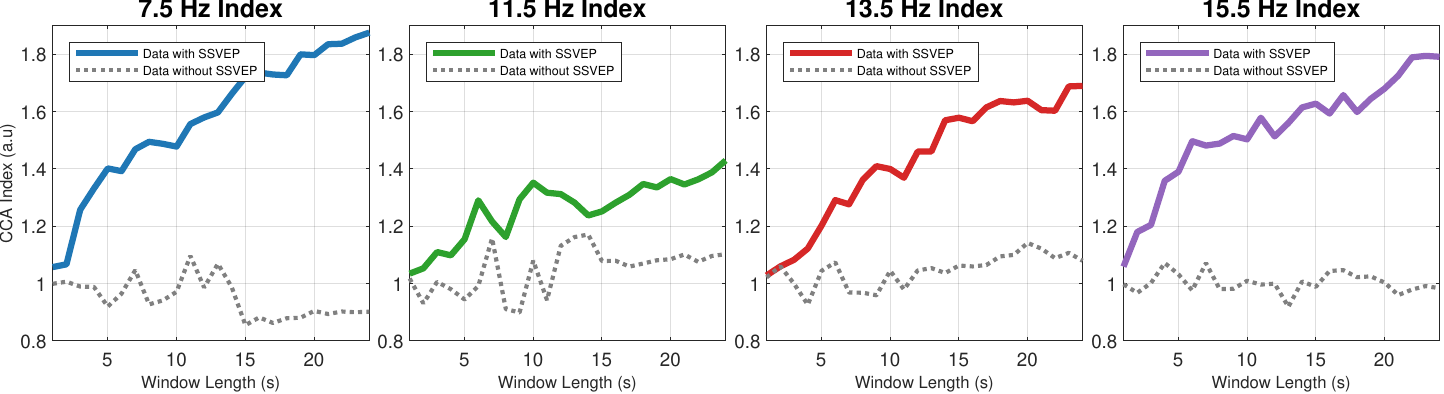}
    }
    
    \vspace{-0.5em}
    
    \subfloat[][(b) CCA-based frequency response to target and non-target trials]{
        \includegraphics[width=0.9\textwidth,page=1]{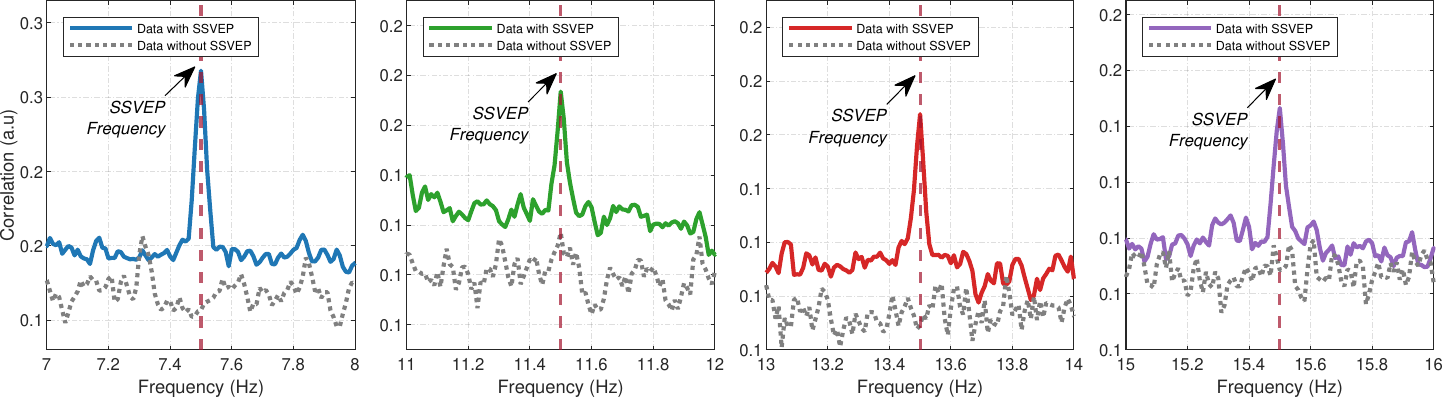}
    }
    
    \caption{SSVEP analysis through NCCA and CCA-based frequency response}
    \label{fig:res_ssvep}
\end{figure*}

\subsection{Power measurements}
We conducted power measurements for each presented form factor, summarized in Table~\ref{table:power_form_factors}.
\reviewnew{
Power measurements were performed using a Power Profiler Kit 2 (PPK2, Nordic Semiconductor)~\cite{PPK_nordic} by powering the device from a battery and monitoring the corresponding battery current and voltage, providing an end-to-end assessment of real operating power. Individual subsystem contributions were obtained by selectively disconnecting the respective subcircuit from the power tree (see Fig.~\ref{fig:MB_PM}) and computing the difference between the total power and the power measured with that subcircuit being powered externally.}
\review{These measurements correspond to streaming-only operation (GAP9 off). Total application power is reported in Table \ref{table:power_edgeAI} and described in Section \ref{in-vivo} (per each deployed application).} The \ac{AFE} operates at \SI{500}{Hz} sampling frequency in high-resolution mode with a \ac{PGA} gain of 12. The PPG sensor is sampling at \SI{100}{Hz} using an LED pulse current of \SI{7.4}{mA}, and the \review{accelerometer} runs at \SI{400}{Hz} in low-power mode.
\begin{table}[h!]
\begin{center}
\begin{threeparttable}[b]
\caption{Power consumption comparison across form factors \review{in streaming mode}.}
\label{table:power_form_factors}
\scriptsize{
\begin{tabular}
{
p{0.9in}>{\centering\arraybackslash}p{0.5in}>{\centering\arraybackslash}p{0.4in}>{\centering\arraybackslash}p{0.6in}}
\toprule[0.20em]
\textbf{Power domain} & \textbf{Headband [mW]} & \textbf{Sleeve [mW]} & \textbf{Chestband [mW]}\\
\midrule
nRF & 8.1 & 8.0 & 2.5\\
ADS digital & 1.9 & 1.9 & 0.7\\
ADS analog & 16.8 & 16.8 & 2.9\\
PPG & 3.1 & - & 3.1\\
Active electrodes & 2.8 & - & -\\
\review{ACC} & 0.06 & 0.06 & 0.06\\
\midrule
\textbf{Total power\footnotemark[1] } & \textbf{32.8} & \textbf{26.7} & \textbf{9.3}\\
\midrule
\textbf{Battery life [h]\reviewnew{\footnotemark[2]}} & \textbf{16.9} & \textbf{20.8} & \textbf{59.7}\\
\bottomrule[0.20em]
\end{tabular}}
\begin{tablenotes}\footnotesize
\item[1] \review{GAP9 power domain is turned off in streaming mode.}
\item[\reviewnew{\textsuperscript{2}}] \reviewnew{Using a 150~mAh LiPo battery (2.1$\times$2.8$\times$0.4 cm$^3$, 4g).}
\end{tablenotes}
\end{threeparttable}
\end{center}
\end{table}

The nRF power consumption includes wireless communication (battery information exchange, sensor data streaming) and general device control (sensor interfacing, power management). The power consumption of the ADS's analog and digital domains mainly depends on the number of active measurement channels. Active electrodes are only used in the case of the headband, as EEG signals have inherently low amplitudes and are noise-sensitive, and should therefore be buffered directly at the electrode. The \review{accelerometer} only accounts for a minor fraction of total power consumption.
Finally, the power consumption of the GAP9 domain is highly application-specific, based on application-specific computational demands. Two typical use cases illustrate its efficiency:
\begin{enumerate}
    \item Responses or underlying trends of biosignals are often evaluated in the frequency domain, hence the computation of FFTs is a common \ac{DSP} load which can be performed at high energy efficiency on GAP9 at the edge. \cite{frey_2023_BioGAP} reports the computation of FFTs (eight floating point FFTs of size 1024 every \SI{50}{ms}) resulting in an energy efficiency of \SI{16.7}{Mflops/s/mW} when operating GAP9 at a \SI{0.65}{V} core voltage and a \SI{240}{MHz} compute cluster frequency. Performing these \ac{DSP} operation on GAP9, compared to streaming the raw data through \ac{BLE} for offline processing resulted in an increase in system reliability (in case of instable BLE throughput) and lower power consumption (\SI{7}{mW} vs. \SI{12}{mW}).
    \item A second commonly used application is to perform inference at the edge with a lightweight \ac{NN} utilizing the NE16 hardware accelerator in GAP9 as reported in \cite{frey_wearable_2024}. The inference of a \ac{CNN} with a size of 21.1k parameters was performed making use of the NE16, a highly energy efficient hardware accelerator included in GAP9, resulting in an energy per inference of only \SI{0.36}{mJ} and an inference time of \SI{10.17}{ms}. The total power consumption of the system (including data acquisition, edge inference, and wireless connection) resulted in a power consumption of \SI{19.6}{mW}.
\end{enumerate}
Overall, the total power consumption measured for the headband, sleeve, and chestband form factors is \reviewnew{\SI{32.8}{mW}}, \SI{26.7}{mW}, and \SI{9.3}{mW}, respectively, when streaming the measured data wirelessly. These values enable continuous operation times of \SI{16.9}{h} (headband), \SI{20.8}{h} (sleeve), and \SI{59.7}{h} (chestband) when \lu{powered by} a compact \SI{150}{mAh} battery.

\section{In-vivo characterization and case studies}
\label{in-vivo}

All participants provided informed consent prior to their participation. All the experimental procedures followed the principles outlined in the Helsinki Declaration of 1975, revised in 2000.

\subsection{EEG Headband for Visual Evoked Potentials} 
\ac{SSVEP} is commonly employed as a control paradigm in \ac{BCI}s. It refers to a frequency- and phase-locked EEG response elicited by repetitive visual stimuli \cite{zhu2010survey}. We assessed the headband's performance \review{over a single subject (male, 35 years old), } positioned approximately \SI{60}{cm} in front of a 14-inch computer monitor. Visual stimuli consisted of sinusoidal on-off patterns with 100\% contrast, presented sequentially on the screen. The stimuli frequencies included \SI{7.5}{Hz}, \SI{11.5}{Hz}, \SI{13.5}{Hz}, and \SI{15.5}{Hz}, each delivered in random order across three repetitions. Each frequency stimulus lasted 25 seconds, interleaved with a 10-second rest period to minimize visual fatigue. The EEG responses were analyzed using the \ac{NCCA} method \cite{kartsch2022efficient}, an extension of \ac{CCA}. \ac{NCCA} calculates the ratio of the \ac{CCA} response at a target frequency relative to the average response at two adjacent side frequencies, consistently set at $\pm$0.2 Hz from the target frequency in this study. The \ac{NCCA} scores were computed for each data segment from individual trials, with results averaged across all trials. \review{All 16 EEG channels available were used (electrode placement detailed in Section \ref{systemDesing:EEGHeadband}). All \ac{NCCA} processing is performed offline.}
Fig.~\ref{fig:res_ssvep} (a) illustrates the \ac{NCCA} results for each target frequency (colored continuous lines) using increasingly larger \ac{CCA} evaluation windows. Additionally, \ac{NCCA} calculations for rest periods (grey dashed lines) serve as a baseline to verify the proper functioning of the \ac{NCCA} algorithm. For all frequencies, a \SI{3}{s} window was found adequate to reliably detect the \ac{SSVEP} response with sufficient confidence\footnote{Empirical observations suggest that \ac{NCCA} values exceeding 1.1 are significantly indicative of the presence of \ac{SSVEP} EEG responses}.
Fig.~\ref{fig:res_ssvep} (b) shows the frequency response computed via the \ac{CCA} algorithm. Each frequency displays distinct power peaks, consistently higher than those observed at adjacent frequencies and during rest intervals.

\subsection{EMG sleeve for gesture recognition} 
\label{sect:emg_use_case}

\review{We validate the EMG sleeve with a multi-subject protocol based on} the movement sequence introduced by Tanzarella et al.\cite{Tanzarella2024}, investigating the synergistic interaction of arm and forearm muscles during functional tasks to enhance hand posture prediction. \review{This task serves as a representative example of a computationally intensive application that necessitates the use of the GAP9 processor for efficient on-device execution.}

\review{\textit{Experimental protocol.} Experiments involved three healthy volunteers (2 male, 1 female, between 25-36 years old). The subject sits comfortably on a chair, and the protocol starts with a rest position (3 seconds), with the elbow bent at $90^\circ$, the wrist aligned with the forearm, and the hand relaxed. The GUI displays visual and audio cues to guide the subject in executing the following movements: reaching and grasping a cylindrical bottle filled with $\approx 2$kg of water placed on a table (the movement lasts 3 seconds), lifting it (3 seconds), placing it back (3 seconds), returning to the rest position (3 seconds), and maintaining rest (3 seconds). Each task lasts a total of 18 seconds.}
\review{Data is collected in 10 sessions. Each session consists of 10 repetitions of the task. Between each repetition, the subject rests for 5 seconds. Between sessions, the subject is given at least 120 seconds to recover, preventing muscle fatigue.}

\begin{figure}[htb]
    \centering
    \includegraphics[width=1\columnwidth, trim={0cm 0cm 0cm 0cm},clip]{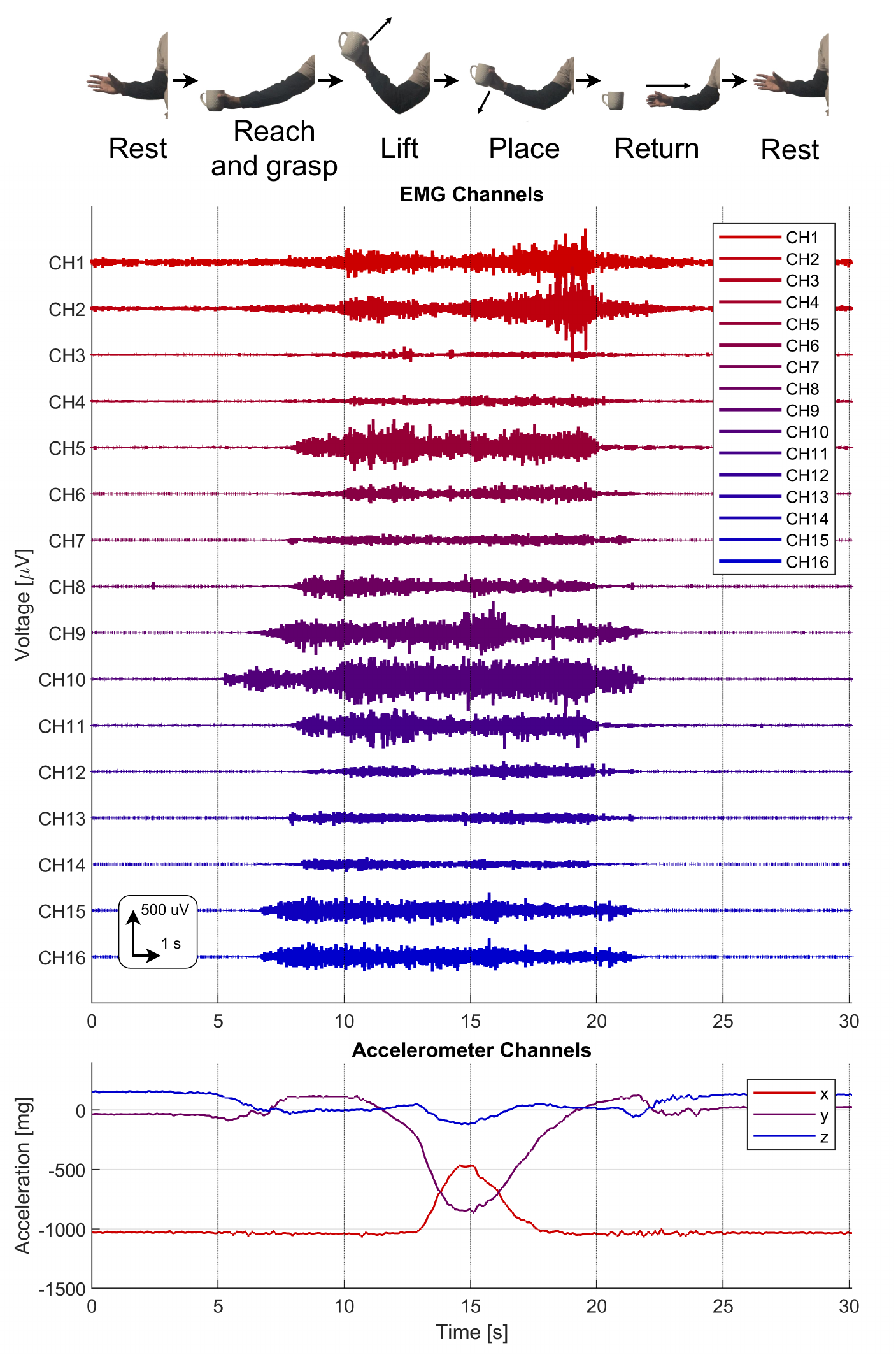}
    
    \caption{EMG sleeve measurement performing a sequence of movements involving different muscle groups}
    \label{fig:res_EMG_Sleeve}
\end{figure}

\review{One example of} recorded EMG signals corresponding to this sequence are illustrated in Fig.~\ref{fig:res_EMG_Sleeve}. 
The acceleration data, shown at the bottom of the figure, tracks limb dynamics and clearly highlights the lifting and placement actions around the 15-second mark.

\review{\textit{Data processing.} EMG data are filtered with an IIR high-pass filter (20 Hz, Butterworth, 4th order), followed by notch filter (50
Hz, Butterworth, 2nd order).
Acceleration data are filtered with an IIR low-pas filter (10 Hz, Butterworth, 2nd order).
Data is then divided into non-overlapping windows of 200 ms. 
}

\review{
\textit{Evaluation procedure.}
We perform a 5-Fold Cross-Validation (CV) with a subject-specific approach. Each fold contains data from all
acquisition sessions (and repetitions).
Data is balanced to have an equal number of labels per class.
}

\review{
\textit{Network architecture.}
We designed an EMG + accelerometer fusion model based on two parallel convolutional branches that independently process EMG and jerk (derivative of the acceleration data). Each branch uses three convolutional blocks (Conv + BatchNorm + ReLU + Dropout + MaxPool). The EMG branch uses temporal kernels of size 10 ($\approx$20 ms) and performs both temporal and spatial downsampling, while the jerk branch uses larger temporal kernels (25 samples, $\approx$ 50 ms) with purely temporal pooling. Kernel sizes for both modalities were selected based on an ablation study. The extracted features from both streams are concatenated and fed to a joint fully connected head (252 $\rightarrow$ 126 $\rightarrow$ 5\reviewnew{)} for movement phase classification. This late-fusion design exploits complementary muscle and motion cues, yielding higher accuracy than single-modality models (not shown).
}

\review{
\textit{Network training.} For all experiments, the models were trained using the Adam optimizer (learning rate = $10^{-3}$, weight decay = $10^{-4}$) with a cross-entropy loss function for up to 50 epochs, applying early stopping when validation loss failed to improve for 5 consecutive epochs.
}

\review{
\textit{Classification results.} We achieved an average 5-class classification accuracy of 79.9\% $\pm$ 5.7\%.  
These results are comparable to similar studies from the literature \cite{liarokapis2013learning, rojas2012high}, which however do not use fully-wearable edgeAI-capable platforms.
}

\review{
\textit{Edge deployment.} The trained network was deployed on the GAP9 \ac{SoC} using 8-bit integer post-training quantization, which significantly reduces memory footprint and computational complexity. The final model comprises 34772 parameters, corresponding to a storage size of about 34.8~kB in its quantized form. NNTool was then used for deployment on the target platform following the same approach as in \cite{frey2024wearable}.}

\review{
Table \ref{tab:deployment} shows the achieved performance metrics, which demonstrate an inference time lower than 5 ms and energy consumption as low as 79.6~\textmu J per inference. 
Table \ref{table:power_edgeAI} (right column) shows the power repartition across the main components of BioGAP-Ultra when doing edgeAI inference. Compared to raw data streaming (Tab.~\ref{table:power_form_factors}) the edgeAI approach enables to further reduce power consumption, while guaranteeing lower latency and improved privacy, as raw data do not require to be transmitted wirelessly.
\begin{table}[htb]
\renewcommand{\arraystretch}{0.95}
  \centering
  \caption{Arm posture prediction CNN deployment performance on GAP9}\label{tab:deployment}
  \vspace{-0.1cm}
  {
    \footnotesize
    \begin{tabular}{lp{2cm}p{2cm}}
      \toprule
      MACs & \multicolumn{2}{c}{539,581}\\
      \cmidrule(r){2-3}
      Frequency [MHz] & 370 & 240 \\
      \cmidrule(r){1-1} \cmidrule(r){2-2} \cmidrule(r){3-3}
      Time/inference [ms]   & 2.89 & 4.42 \\
      Cycles/inference & 1,060,530 & 1,060,530 \\
      Operations/cycle & 0.509 & 0.509 \\
      Energy/inference [$\mu J$] & 114.9 & 79.6 \\
      \bottomrule
    \end{tabular}
  }
\end{table}

\begin{table}[h!]
\centering
\begin{threeparttable}

\caption{\review{Power consumption for two representative edge processing workloads.}}
\label{table:power_edgeAI}
\scriptsize
\review{
\begin{tabular}{p{1.1in} >{\centering\arraybackslash}p{0.7in} >{\centering\arraybackslash}p{0.7in}}
\toprule
\textbf{Power domain} & \textbf{Pan–Tompkins PAT [mW]} & \textbf{Arm posture [mW]} \\
\midrule
nRF & 1.9 & 0.8 \\
ADS digital & 0.7 & 1.9 \\
ADS analog & 2.9 & 16.8 \\
PPG & 3.1 & -- \\
Active electrodes & -- & -- \\
\review{ACC} & 0.06 & 0.06 \\
GAP9 & -- & 4.05 \\
\midrule
\textbf{Total power} & \textbf{8.6} & \textbf{23.6} \\
\midrule
\textbf{\reviewnew{Corresponding power in streaming mode}\reviewnew{\footnotemark[1]}} & \textbf{\reviewnew{9.3}} & \textbf{\reviewnew{26.7}} \\
\midrule
\textbf{Battery life [h]\reviewnew{\footnotemark[2]}} & \textbf{64.2} & \textbf{23.6} \\
\bottomrule
\end{tabular}
}
\vspace{0.1cm}

\begin{tablenotes}
\footnotesize
\item[\reviewnew{\textsuperscript{1}}] \reviewnew{See Table \ref{table:power_form_factors} for a detailed power breakdown in streaming mode.}
\item[\reviewnew{\textsuperscript{2}}] \reviewnew{Using a 150~mAh LiPo battery (2.1$\times$2.8$\times$0.4 cm$^3$, 4g).}
\end{tablenotes}
\end{threeparttable}
\end{table}

}

\subsection{ECG band and PPG clip for PAT}

\review{
We validate the ECG chestband and PPG clip by implementing an online \ac{PAT} extraction task. We choose to measure the peak-based PAT, instead of foot-based PAT \cite{block2020conventional}, because the PPG peak can be identified more robustly in noisy wearable recordings. This task serves as a representative example of a use case with low computational requirements (compared to the more involved computations required for the EMG gesture classification), which can be executed directly on the nRF5340 MCU.
}

\review{\textit{Experimental protocol.}}
\review{Experiments involved three healthy subjects (2 male, 1 female, aged between 25--36 years old). Each subject sat comfortably on a chair while wearing the ECG chestband, with the PPG sensor attached to the right earlobe using a clip. Additionally, to further validate the earlobe measurements, we also conducted experiments when measuring PPG from the right index finger. Measurements were recorded for \SI{60}{s} per location under resting conditions to ensure stable cardiovascular signals.}

\review{\textit{Data processing and PAT extraction.}}
\review{Both ECG and PPG signals were processed on the nRF5340 microcontroller using the CMSIS biquad implementation \cite{cmsis_dsp} with a 10th-order IIR Butterworth band-pass filter, with cutoff frequencies of \SI{0.5}{Hz} to \SI{30}{Hz} for ECG and \SI{0.5}{Hz} to \SI{15}{Hz} for PPG, isolating the physiological frequency components relevant for \ac{PAT} estimation. Peak detection for both waveforms was performed using a real-time implementation of the Pan–Tompkins algorithm running on the nRF5340 microcontroller, following the open-source design in \cite{PanTompkinsOpenSource}. For each cardiac cycle, the time delay between the ECG R-peak and the subsequent PPG systolic peak was computed, yielding the peak-based \ac{PAT}.}

\begin{figure}[htb]
    \centering
    \includegraphics[width=1\columnwidth, trim={0cm 0cm 0cm 0cm},clip]{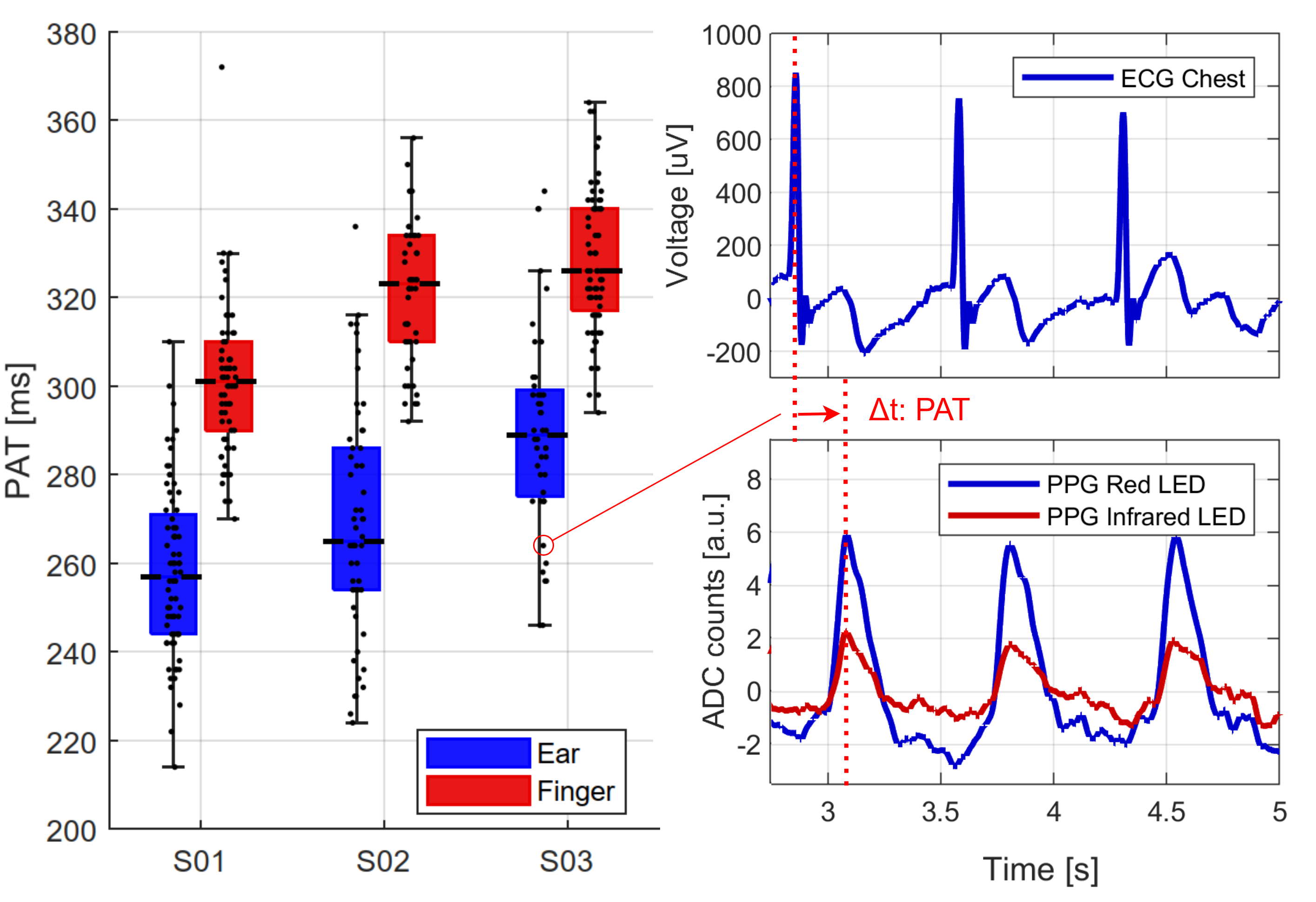}
    
    \caption{\review{Pan-Tompkins PAT times per subject and location with inset of ECG-PPG waveform.}}
    \label{fig:res_ecg_band}
\end{figure}

\review{\textit{Results.}} 
\review{
The distribution of extracted peak-based \ac{PAT} values across subjects and measurement sites is shown in Fig.~\ref{fig:res_ecg_band}. The inset also depicts an example of synchronized ECG and PPG waveforms, with the corresponding \ac{PAT} interval indicated, representing the propagation time of the pulse wave from the heart to the peripheral measurement site. The difference between earlobe-based and finger-based PAT is approximately 37 to 58~ms for all subjects. The peak-based PAT is slightly longer than the foot-based PAT definition adopted in several studies \cite{block2020conventional}, and our measured values fall within the expected range once accounting for the additional waveform rise time inherent to the peak-based measurement. 
}

\review{\textit{Edge deployment.}}
\review{The algorithm operated in real time, processing 100-sample chunks every 200~ms, with a moving-window integrator of 150~ms for ECG and 200~ms for PPG. Hard and soft refractory periods of 200~ms/360~ms for ECG and 250~ms/600~ms for PPG prevented false detection. Peak extraction for a 200~ms window of ECG and PPG data was completed in 20.66~ms, corresponding to an energy per computation chunk of \SI{223.7}{\mu J}.}
\review{The power breakdown as well as the total power consumption is listed in Table~\ref{table:power_edgeAI}.}


\section{Discussion}
\label{sect:discussion}

\mbox{BioGAP\lu{-Ultra}} extends the capabilities of the previous design  presented in \cite{frey_2023_BioGAP}, significantly improving flexibility, multimodal integration, and usability. This enhanced version features a modular hardware architecture that enables the simultaneous and fully synchronized acquisition of various biosignals via an integrated ExG expansion board alongside customizable board templates tailored for specific application needs. Regarding electrophysiological (ExG) signal capabilities, \mbox{BioGAP\lu{-Ultra}} doubles the channel count and supports flexible configuration options, including single-ended and fully differential setups, effectively addressing a broader range of research and clinical requirements.
Furthermore, \mbox{BioGAP\lu{-Ultra}} continues to rely on the GAP9 processor, in light of its exceptional energy efficiency in edge-AI applications. The upgraded platform incorporates a new BLE module (nRF53), enhancing data throughput and significantly reducing power consumption compared to its predecessor, which markedly improves real-time data handling and prolongs battery life. \review{Moreover, \mbox{BioGAP\lu{-Ultra}} integrates larger RAM and Flash, enabling deployment of larger models, longer on-device data retention for feature extraction, and resilient buffering during transient BLE dropouts. }
Complementing the hardware advancements, \mbox{BioGAP\lu{-Ultra}} also provides an integrated software suite designed to streamline data collection, facilitate real-time visualization, and enable comprehensive data analysis, thus simplifying experimental procedures and data management considerably.
\review{Additionally, we reported the platform integration in multiple form factors together with related edgeAI demonstrations.}
Lastly, by adopting a fully open-source model, \mbox{BioGAP\lu{-Ultra}} promotes broader accessibility and fosters innovation, empowering the research community to conveniently adapt, enhance, and further develop this sophisticated multimodal biosensing platform.

\subsection{Comparison to state-of-the-art}
\label{sect:soa_comparison}
Compared to existing state-of-the-art multimodal wearable platforms, \mbox{BioGAP\lu{-Ultra}} uniquely combines modularity, synchronized multimodal acquisition, and on-device AI capabilities within a low-power, fully open-source framework. As highlighted in Table~\ref{table:soa_comparison}, most platforms lack embedded AI support; only \cite{najafi_versasens_2024} integrates dedicated edge-AI resources. However, VersaSens exhibits limited form factor adaptability and incurs significantly higher power consumption (>\qty{130}{\milli\watt}). Conversely, energy-efficient platforms such as \cite{song_769_2019} operate at sub-milliwatt levels (~\qty{0.77}{\milli\watt}) but support only a restricted set of modalities and do not include on-device inference. \mbox{BioGAP\lu{-Ultra}} addresses these limitations by integrating the GAP9 SoC, delivering up to \qty{32.2}{GMAC/s} for real-time inference while supporting full-stack acquisition of low-SNR signals, including EEG, EMG, ECG, and PPG. Although some platforms support a broader range of physiological parameters, they often avoid EMG or EEG acquisition, core biosignals for medical applications, due to stringent hardware constraints. In contrast, \mbox{BioGAP\lu{-Ultra}} provides hardware-synchronized ExG acquisition while allowing extending the number of modalities over expansion boards without compromising compactness. Furthermore, it incorporates the latest BLE 5.4 SoC (nRF53), enabling high-throughput, energy-efficient wireless communication. Notably, it is the only platform among those compared to offer a comprehensive software suite for real-time data visualization, configuration, and storage, thereby streamlining experimental workflows for researchers.


These results underscore the versatility and effectiveness of \mbox{BioGAP\lu{-Ultra}} as a next-generation wearable biosensing platform. By integrating a scalable hardware-software stack, low-noise ExG acquisition, synchronized multimodal sensing, and energy-efficient edge-AI execution within a compact, modular architecture, \mbox{BioGAP\lu{-Ultra}} directly addresses the key limitations observed in prior systems and lays the groundwork for more responsive, robust, and intelligent wearable health systems.

\subsection{Broader applicability}\label{sect:study_cases}
We conclude the discussion by describing, for each of the form factors, applications and edge AI tasks that \review{could} benefit from the new multimodal, wearable sensing platform. 


\subsubsection{Headband}
The measurement of EEG and PPG signals in wearable headbands is crucial for non-obtrusive neurological monitoring applications, such as seizure detection or drowsiness monitoring, where simultaneous multimodal sensing greatly enhances detection accuracy and reliability.

Ingolfsson et al. \cite{ingolfsson2024brainfusenet} showed the effective integration of \ac{EEG} and \ac{PPG} signals collected through wearable devices for seizure detection, demonstrating the successful detection 92\% of seizure events, while reducing false positives to less than 1 per day. Implemented on the GAP9 platform, their approach achieved an energy efficiency of 21.43 GMAC/s/W and an energy consumption per inference of just 0.11 mJ at high performance (412.54 MMAC/s). This highlights the potential of directly running these algorithms onboard GAP9 integrated in our headband, enabling a fully integrated wearable solution for simultaneous EEG and PPG monitoring for non-obtrusive seizure detection.
The increased number of EEG channels (16) of our headband, compared to the prototype of \cite{wang2023enhancing}, also holds promise for increased accuracy across various BCI tasks, \reviewnew{such as motor-movement \cite{wang2023enhancing}, speech imagery \cite{ingolfsson2025wearable}, and drowsiness detection \cite{frey2024wearable}}.



\subsubsection{Sleeve}
\ac{EMG}-based gesture recognition, prosthetic control, and tracking complex motor tasks have shown high potential, but current studies lack the full integration of measurement systems and processing capabilities in a fully wearable form factor.
A recent study by Tanzarella et al. \cite{tanzarella_arm_2024} explored how combining \ac{EMG} signals from both upper arm and forearm muscles could enhance predictions of hand posture during reaching, grasping, and manipulation tasks. Participants performed several natural grasping movements, while muscle activity was recorded using bipolar electrodes on upper arm muscles (including deltoids, biceps, and triceps) and surface EMG grids on forearm flexor and extensor muscle groups. Concurrently, hand joint angles were tracked using a motion-capture system.
The analysis of muscle and postural synergies demonstrated that predictions of hand posture improved significantly when information from both the upper arm and forearm was combined, as opposed to relying on one muscle group alone, demonstrating the benefit of combined forearm and upper arm EMG measurements. The study’s primary limitation, the use of a laboratory-based EMG setup, which hinders the adoption outside a lab environment, can be effectively overcome with our proposed wearable EMG sleeve that allows for combined forearm and upper arm EMG measurements in a fully wearable setup.

In another work, Tacca et al. \cite{tacca_wearable_2024} have shown a wearable high-density EMG sleeve to classify complex hand gestures and estimate continuous joint angles using machine learning models, specifically neural networks. EMG signals were processed using \ac{RMS} features extracted from 100 ms bins. The neural network achieved high performance, classifying 37 sequential hand gestures with approximately 93\% accuracy. This demonstrates significant potential for real-life applications if integrated with our wearable measurement system and the onboard computing capabilities provided by GAP9, which efficiently supports neural network computations.


\subsubsection{Chestband}
Cuffless \ac{BPM} methods are essential as they allow continuous, unobtrusive monitoring without the discomfort and limitations of traditional cuff-based systems.

Yoon et al. \cite{yoon_non-constrained_2009} introduced a non-constrained \ac{BPM} method using ECG and PPG signals. Their approach extracted key indices such as the pulse transit time from the R-peak ECG to the maximum first derivative of the PPG signal, the pulse transit time to waveform onset, the systolic upstroke time, the diastolic time, and the width at two-thirds of the amplitude time of the PPG. 

Other recent works based on \ac{ECG} - \ac{PPG} fusion and machine learning have shown robust performance in \ac{BPM} using deep learning models \cite{long_bpnet_2023} or feature extraction followed by classifiers \cite{li_noninvasive_2024}. These works demonstrate the potential of a wearable measurement of combined ECG and PPG, including onboard processing capabilities, which is enabled by our proposed platform.



\subsubsection{Foundation models on wearables}

\reviewnew{Beyond conventional ML workloads, the architectural choice of including the GAP9 chip and significant external memory is driven by the emerging trend toward foundation models (FM) for biosignals, which leverage large-scale pretraining on unlabelled data to improve generalization. In particular, recent works demonstrated the feasibility of deploying sub-10M parameter FMs directly on GAP9 \cite{dimofte2025cerebro,Ingolfsson_2025_phd_thesis}, using the limited onboard L2 memory as a streaming buffer while storing weights and activations in an external L3 memory.
The use cases reported in this work can also be implemented as downstream tasks for such FMs, enabling more robust classification and regression directly on BioGAP-Ultra.}


\subsection{Limitations}
\label{sect:limitations}

\review{The current implementation of BioGAP-Ultra presents the following limitations, which will be addressed as part of future work.}

\begin{enumerate}
    \item  \review{\textit{On-board storage:} data streaming presently requires an external device; however, the modular design allows straightforward integration of an SD-card or flash-based expansion for fully autonomous operation. The additional modules can be built using small-footprint parts with little impact on device size.}
    \item  \review{\textit{Wired interconnection:} in the present validation setup, the chestband and earlobe PPG sensor are connected via cable to ensure synchronization for PAT measurements. Future revisions will integrate low-power wireless synchronization between nodes.}
    \item \review{\textit{Montage coverage:} the fixed headband/sleeve electrode locations limit applications scope,  addressable by swapping/adding form factors to match task needs.}
    \item \review{\textit{Population size:} validation was performed on a limited number of subjects. Larger cohorts and multi-session evaluations are planned as part of ongoing work.}
\end{enumerate}

\section{Conclusion}\label{sec:conclusion}

In this paper, we introduced \textsc{BioGAP\lu{-Ultra}}, a modular platform designed for wearable multimodal biosignal acquisition and efficient edge-AI processing. We described the system design, which features a Mainboard with a dual-SoC architecture utilizing GAP9 and nRF5340 for power management, system control, wireless connectivity, and embedded AI capabilities. The Mainboard can be flexibly paired with different sensing boards to expand its sensing capabilities for multimodal data acquisition (EEG, EMG, ECG, PPG, \review{ACC}). 

We demonstrated the versatility of \textsc{BioGAP\lu{-Ultra}} by customizing and integrating it into various wearable form factors: an EEG-PPG headband, an EMG \review{sleeve}, and an ECG-PPG chestband. We characterized system power consumption for each form factor, measuring \SI{32.8}{mW}, \SI{26.7}{mW}, and \SI{9.3}{mW} \review{in streaming mode}, respectively, enabling continuous operation for up to \SI{59.7}{h} using a compact \SI{150}{mAh} battery. We \review{demonstrate} the platform's functionality \review{through deployment and characterization of two edgeAI applications: ECG-PPG chestband-based \ac{PAT} estimation within a power envelope of \SI{8.6}{mW} and EMG+ACC sleeve-based classification of the different phases of a continuous reach-and-grasp arm movement, achieving an accuracy of 79.9\% $\pm$ 5.7\% at \SI{23.6}{mW}}, and provided insight into practical, real-world applications suitable for each configuration.

Overall, \textsc{BioGAP\lu{-Ultra}} advances the \ac{SoA} in wearable multimodal sensing by combining modularity, low-power operation, and embedded intelligence within flexible and user-friendly form factors. The complete hardware, firmware, and software stack is open-sourced.  Future work will explore the integration of additional sensing modalities, such as an ultra-low power wearable ultrasound probe like WULPUS \cite{frey2022wulpus} \lu{or functional near-infrared spectroscopy (fNIRS) \cite{lee2025review, rahimpour2025wearable}}, to further extend the capabilities and application scope of the presented platform.  \review{Moreover, while in its present form our platform requires the Android device for data reception and storage, future work will also integrate microSD-based onboard storage for fully autonomous data collection capabilities. \reviewnew{Additionally, we will extend the platform towards wireless synchronization of multiple sensing nodes (e.g., headband, armband, chestband, earclip) to reduce cables and enable concurrent multimodal acquisition in larger cohorts.}}

\luca{All hardware and software design files are released with a permissive open source license: \href{https://github.com/pulp-bio/BioGAP}{https://github.com/pulp-bio/BioGAP}}

\section*{Acknowledgment}
\vspace{-0.1cm}
We thank P. Wiese, L. Sivakolunthu, A. Blanco Fontao, and H. Gisler (ETH Zürich) for technical support.

\bibliographystyle{IEEEtran}
\bibliography{bib/main}

\end{document}